\documentclass[10pt,a4paper,twoside]{article}
\usepackage[utf8]{inputenc}
\usepackage[T1]{fontenc}
\usepackage{natbib}
\usepackage[french]{babel}

\usepackage{amssymb}
\usepackage{amsthm}
\usepackage{amsmath}
\usepackage{amsfonts}

\usepackage{dsfont}
\usepackage{color}
\usepackage{latexsym}
\usepackage{graphicx}
\usepackage[toc,page]{appendix}
\usepackage{pgf}
\usepackage[margin=3.5cm]{geometry}
\usepackage{hyperref}
\usepackage{authblk}
\usepackage[nottoc, notlof, notlot]{tocbibind}
\usepackage{varwidth}

\usepackage{fancyhdr}
\newcommand{\E}{\mathbb{E}}
\newcommand{\R}{\mathbb{R}}

\renewcommand{\P}{\textup{P}}

\renewcommand{\leq}{\leqslant}
\newcommand{\ctt}[1]{\texttt{#1}}
\newcommand{\argmin}{\mathop{\mbox{argmin}}}

\definecolor{nathgrey}{HTML}{D4C8D4}

\newcommand{\amax}[1]{\underset{#1}{\textup{argmax} \,}}

\newtheorem{definition}{Définition}

\newcommand{\Tf}{\widetilde{T}}

\title{\textbf{{\huge Arbres CART et Forêts aléatoires} \\
Importance et sélection de variables}}

\author[1]{Robin Genuer}
\author[2]{Jean-Michel Poggi}

\affil[1]{\small{ISPED, Univ. Bordeaux, INSERM U-1219 \& INRIA, SISTM, France \\ robin.genuer@u-bordeaux.fr}}
\affil[2]{\small{
{\protect\begin{varwidth}[t]{\linewidth}\protect\centering LMO, Univ. Paris-Sud Orsay \& Univ. Paris Descartes, France, \par jean-michel.poggi@math.u-psud.fr\protect\end{varwidth}}}}

\begin{document}
\maketitle
\thispagestyle{empty}

\vspace*{1cm}

\begin{abstract}
Deux des algorithmes proposés 
par Leo Breiman : les arbres CART (pour Classification And Regression Trees)
introduits dans la première moitié des années 80
et les forêts aléatoires apparues, quant à elles, au début des années 2000, font l'objet de cet article.
L'objectif est de proposer sur chacun des thèmes abordés, un exposé, une garantie théorique, un exemple
et signaler variantes et extensions.
Après un préambule, l'introduction rappelle les
objectifs des problèmes de classification et de régression avant de retracer 
quelques prédécesseurs
des forêts aléatoires. Ensuite, une section est consacrée aux arbres CART 
puis les forêts aléatoires sont présentées. 
Ensuite, une procédure de sélection de variables basée sur la quantification de 
l'importance des variables est proposée. Enfin l'adaptation des forêts 
aléatoires au contexte du Big Data est esquissée. \\

\noindent
\textbf{Mots clés}: CART, Forêts aléatoires, Importance des variables, Sélection de variables, Big data.  \\

\bigskip

\begin{center} \textbf{Abstract} \end{center}

Two algorithms proposed
by Leo Breiman: CART trees (Classification And Regression Trees for)
introduced in the first half of the 80s
and random forests emerged, meanwhile, in the early 2000s, are the subject of this article.
The goal is to provide each of the topics, a presentation, a theoretical guarantee, an example
and some variants and extensions.
After a preamble, introduction recalls
objectives of classification and regression problems before retracing
some predecessors of the Random Forests. Then, a section is devoted to CART trees
then random forests are presented.
Then, a variable selection procedure based on permutation variable
importance is proposed. Finally the adaptation of 
random forests to the Big Data context is sketched.  \\


\end{abstract}

\newpage

\tableofcontents

\newpage

\mbox{}

\newpage

\section{Préambule}\label{Préambule}

Comme l'indique son titre, cet article traite de deux des algorithmes 
proposés 
par Leo Breiman : les arbres CART (pour Classification And Regression Trees)
introduits dans la première moitié des années 80
et les forêts aléatoires apparues, quant à elles, au début des années 2000, 
marquant 
ainsi une période d'intense évolution conjointe de la statistique et de ce que 
l'on appelle aujourd'hui l'apprentissage statistique. La trajectoire des intérêts de Leo Breiman, 
dont la biographie scientifique est esquissée dans 
l'intéressante conversation \cite{Olshen2001} et
dans \cite{cutler2010remembering}, constitue
sans nul doute un trait magistral de ces disciplines. Il a en effet, après avoir 
développé son activité 
dans le domaine des probabilités sous un angle très
proche des mathématiques pures, marqué de son empreinte la statistique 
appliquée
et l'apprentissage.

Les arbres de décision donnent lieu à des méthodes
très versatiles permettant de traiter semblablement le cas de la régression, de 
la classification bi-classe ou multi-classe ou encore de mélanger des variables 
explicatives quantitatives et qualitatives.  CART est d'une certaine manière la 
première pierre de l'édifice des méthodes d'arbres qui, en général, héritent
de ces propriétés. En effet, bien que connue depuis les années 60, la méthode des 
arbres de décision souffrait de fortes critiques justifiées et CART 
leur offre un cadre conceptuel de type sélection 
de modèles, qui leur confère ainsi à la fois une large applicabilité, 
une facilité d'interprétation et des garanties théoriques. 

Mais l'un des défauts majeurs, intrinsèque, demeure : l'instabilité. Le remède, original 
et très profondément
statistique, consiste à exploiter la variabilité naturelle des méthodes 
d'estimation en conjuguant deux 
mécanismes fondamentaux : la
perturbation aléatoire des arbres et la combinaison d'un ensemble d'arbres
plutôt que la sélection de l'un d'entre eux. Plusieurs algorithmes ont été proposés,
la plupart par Breiman, comme le Bagging 
(\cite{Breiman96}) et diverses variantes de Arcing (\cite{Breiman98}) 
mais pas tous, en particulier le célèbre 
Adaboost (\cite{FREUND1997119}) qui fut développé sur des bases conceptuelles 
assez différentes (théorie des jeux). Cet
ensemble de propositions sont surpassées, au moins
sur le plan expérimental, par les forêts aléatoires (abrégées parfois RF pour Random Forests dans la suite).

En effet, les forêts aléatoires sont une méthode de statistique non-paramétrique aux
performances exceptionnelles très tôt remarquées et sans cesse confirmées depuis. 
Même si ce type d'exercice est un peu 
artificiel, on peut le saisir au travers des multiples évaluations massives menées
périodiquement dans les revues appliquées de Machine Learning (comme par 
exemple \textit{Journal of Machine Learning Research}  ou 
encore \textit{Pattern Recognition Letters}) et dans lesquelles 
elles ressortent systématiquement dans les 2 ou 3 meilleures. La dernière et très remarquée peut être trouvée dans le 
papier de \cite{fernandez2014we}  qui couronne les RF, alors que, moins d'une dizaine d'années auparavant, le papier 
\cite{wu2008top} mentionne CART mais pas encore les forêts aléatoires. Introduites 
par \cite{Breiman01}, elles sont depuis de plus en plus utilisées pour traiter de
nombreuses données réelles dans des domaines d'application
variés, citons par exemple l'étude des biopuces (\cite{Diaz06}), 
l'écologie (\cite{Prasad06}), la prévision de la pollution 
(\cite{ghattas1999previsions}) ou encore la génomique (\cite{Goldstein10} et \cite{Boulesteix12}), et
pour une revue plus large, voir \cite{Verikas11}.

Au delà des performances et du caractère automatique de la méthode avec très peu de paramètres à régler,
l'un des aspects les plus importants sur le plan appliqué est sans 
nul doute la quantification de l'importance des variables. Cette notion peu examinée 
par les statisticiens (voir par exemple le papier de synthèse  
\cite{gromping2015variable} en régression) trouve dans le cadre des forêts aléatoires une définition 
commode, facile à évaluer et s'étendant naturellement au cas des groupes de variables 
(voir \cite{gregorutti2015grouped}).

\medskip

Le plan de ce papier est le suivant. Ce préambule se poursuit par la section 
\ref{introduction} qui rappelle les
objectifs des problèmes de classification et de régression avant de retracer 
quelques prédécesseurs
des forêts aléatoires. La section \ref{arbres-cart} est consacré aux arbres CART 
suivie de la section 
\ref{des-arbres-aux-forets-aleatoires} présentant les forêts aléatoires. 
Ensuite, la section 
\ref{forets-aleatoires-et-selection-de-variables} propose
une procédure de sélection de variables basée sur la quantification de 
l'importance des variables. Enfin la
section  \ref{forets-aleatoires-et-big-data} esquisse l'adaptation des forêts 
aléatoires au contexte du Big Data.

\section{Introduction}\label{introduction}

Le problème de base consiste à s'intéresser au couple aléatoire $(X,Y)$, où les 
variables explicatives
 $X\in\mathcal{X}$ (typiquement $\mathcal{X}=\mathbb{R}^p$  mais on peut aussi considérer 
$\mathcal{X}$ de la forme $\mathcal{X}=\mathbb{R}^{p'}\otimes\mathcal{Q}$ pour mélanger variables explicatives quantitatives et qualitatives) sont utilisées
pour expliquer la variable réponse
$Y\in\mathcal{Y}$. L'espace $ \mathcal{Y}=\R $ pour la régression et  $ 
\mathcal{Y}=\{1,\ldots,L\} $ pour la  classification.
Le cadre général est donc celui de l'estimation ou de la prédiction à partir de 
données d'un échantillon
 $\mathcal{L}_n=\{(X_1,Y_1),\ldots,(X_n,Y_n)\}$, les $X_i$ sont appelées entrées et les $Y_i$  sorties, vues 
 comme la réalisation de $n$ 
variables aléatoires i.i.d. de même loi que $(X,Y)$.

\subsection{Objectifs}\label{objectifs}

Nous parlons  d'un problème de prédiction lorsque
la prédiction \(\hat{y}\) doit être la plus proche possible de la vraie
réponse \(y\), associée à \(x\). 
Il existe une autre façon de voir le problème, c'est le point de vue de
l'estimation. Il s'agit dans ce cas d'estimer la fonction (inconnue) qui à
\(X\) associe \(Y\). Bien entendu, ce problème est relié au   
précédent : si nous disposons d'une ``bonne''
estimation du lien entre \(X\) et \(Y\), nous pourrons a fortiori
``bien'' prédire la sortie correspondant à une nouvelle entrée \(x\). Néanmoins, \textit{a contrario}, il
est parfois possible de bien prédire alors que l'estimation de la
fonction qui relie \(X\) et \(Y\) n'est pas très bonne.

Il existe deux principaux cadres en apprentissage statistique : la
régression et la classification, qui diffèrent par la nature
de la sortie \(Y\). 

\subsubsection{Régression}

Le cadre de la régression est celui où la réponse \(Y\) est continue,
typiquement lorsque \(\mathcal{Y} = \mathbb{R}\). Le modèle statistique
s'écrit alors sous la forme suivante :

\begin{align}\label{regression}
 Y=s(X)+\varepsilon
\end{align}

La fonction de régression \(s:\mathcal{X} \to \mathbb{R}\) est inconnue
et nous cherchons à l'estimer à partir des mesures $(X_i,Y_i)$ dont nous disposons
dans l'échantillon \(\mathcal{L}_n\).  Ces mesures sont  
des observations de \(s(X_i)\) bruitées par des variables aléatoires 
\(\varepsilon_i\).
Pour des raisons d'identifiabilité, 
nous supposons que la variable de
bruit \(\varepsilon\) est centrée conditionnellement à \(X\) :
\(\mathrm{E}[\varepsilon|X]=0\). Il existe alors une unique
fonction \(s\) qui satisfait \(s(X)=\mathrm{E}[Y|X]\).

Ce modèle statistique est appelé modèle de régression non-paramétrique
puisqu'essentiellement aucune contrainte \textit{a priori} n'est imposée à la fonction de 
régression \(s\), 
contrairement aux modèles
paramétriques comme par exemple le modèle de régression linéaire. Dans
un tel modèle, on cherche  en effet \(s\) sous la forme d'une combinaison linéaire des 
coordonnées des composantes de \(X\) et
les coefficients de cette combinaison linéaire, appelés les
paramètres du modèle, sont à estimer.

Dans le cadre du modèle (\ref{regression}), nous
introduisons deux mesures de qualité : l'une pour le problème de
prédiction, l'autre pour le problème d'estimation.

\begin{itemize}
    \item Étant donné un prédicteur $\hat{h}$, c'est-à-dire une fonction de 
$\mathcal{X}$ 
    dans $\mathbb{R}$, construite sur l'échantillon d'apprentissage 
$\mathcal{L}_n$. Le but de $\hat{h}$ 
    est de prédire la sortie $y$ associée à une entrée $x$. Nous mesurons la 
qualité de $\hat{h}$ par son 
    erreur de généralisation, définie par :

$$\mathrm{E}[(\hat{h}(X)-Y)^2] \; .$$

   \item Pour le problème d'estimation, nous disposons d'un estimateur $\hat{s}$ 
de la fonction de 
   régression $s$, c'est-à-dire une fonction de $\mathcal{X}$ dans $\mathbb{R}$, 
construite sur 
   l'échantillon d'apprentissage $\mathcal{L}_n$. Le but de $\hat{s}$ est 
d'estimer au mieux la 
   fonction $s$. Nous mesurons la qualité de $\hat{s}$ par son risque, défini 
par :

$$\mathrm{E}[(\hat{s}(X)-s(X))^2] \; .$$

\end{itemize}

Ces deux mesures de qualité dépendent donc du point de vue (prédiction
ou estimation). De plus, comme nous avons supposé que
\(\mathrm{E}[\varepsilon|X]=0\), ces deux mesures satisfont la relation
suivante : pour un prédicteur \(\hat{h}\),

\[\mathrm{E}[(\hat{h}(X)-Y)^2] = \mathrm{E}[(\hat{h}(X)-s(X))^2] + 
\mathrm{E}[\varepsilon^2] \; .\]

Ainsi, en régression, la différence entre prédiction et estimation est
essentiellement une différence de point de vue et de vocabulaire. Comme
nous allons maintenant le voir, ce n'est pas le cas en classification.

\subsubsection{Classification}

En classification (appelée souvent classification supervisée),
la réponse \(Y\) est discrète et désigne la classe (ou le label, l'étiquette de 
la classe) à laquelle appartient
l'entrée \(X\) associée. Ici, \(\mathcal{Y} = \{1,\ldots,L\}\), où \(L\)
désigne le nombre de classes. Nous codons l'ensemble des classes de
façon ordonnée pour faciliter les notations, mais l'ensemble des classes
n'a en général pas de structure,  \(Y\) est nominale (ou catégorielle).

En régression, le but est d'estimer la fonction de régression, qui n'est
autre que l'espérance conditionnelle de \(Y\) sachant \(X\). En
classification, nous ne pouvons pas écrire le modèle sous une forme
équivalente au modèle (\ref{regression}). Mais le but
est maintenant d'estimer les probabilités \textit{a posteriori} définies, pour un
\(x\in\mathcal{X}\) fixé, par :

\[\forall c \in \{1,\ldots,L\} \quad \P(Y=c|X=x)\]

\noindent
c'est-à-dire les probabilités pour \(Y\) d'appartenir à chacune des
classes, conditionnellement à \(X\).

Le fait que nous traitions un échantillon d'observations bruitées implique  
que pour un \(x\) fixé, il n'y a pas forcément une
probabilité \textit{a posteriori}, parmi les $L$, qui soit égale à \(1\) et les autres égales à \(0\).
Donc, pour certaines observations, la classe correspondante à \(x\)
devrait être \(c\), mais se retrouve altérée en \(c'\) dans
l'échantillon. En régression, le bruit vient du fait que nous
n'observons pas exactement \(s(X)\), mais \(s(X)+\varepsilon\), alors qu'en
classification, le bruit provient du fait que certaines étiquettes des classes sont
altérées.

En classification, nous avons également deux mesures de qualité : l'une
pour la prédiction, l'autre pour l'estimation.

\begin{itemize}
    \item Nous mesurons la qualité d'un prédicteur $\hat{h}$ par son erreur de 
généralisation, définie par :

$$\P(\hat{h}(X) \neq Y). $$

\medskip

    \item Le prédicteur qui minimise l'erreur de généralisation est appelé 
prédicteur de Bayes. 
    Ce prédicteur prédit pour un $x$ fixé la quantité suivante : $$\amax{c \in 
\{1,\ldots,L\}} \P(Y=c|X=x).$$
    Bien entendu, ce prédicteur n'est calculable que si l'on connaît la loi du 
couple $(X,Y)$. C'est un 
    estimateur idéal qu'on cherche à approcher.

Notons $\hat{p}(x,c)$ un estimateur de $\P(Y=c|X=x)$, la probabilité \textit{a posteriori}. Une façon d'approcher 
le prédicteur de Bayes est alors de proposer un prédicteur $\hat{h}$ qui prédit, 
pour un $x$ donné, la 
quantité $\amax{c \in \{1,\ldots,L\}} \hat{p}(x,c)$. Nous pouvons alors mesurer 
la capacité du 
prédicteur $\hat{h}$ à bien estimer le prédicteur de Bayes, par exemple, par la 
quantité suivante :

$$\mathrm{E}\left[\sum_{c=1}^L | \, \hat{p}(X,c)- \P(Y=c|X) \, | \right] .$$
\end{itemize}

\medskip

Dans ce cadre, il n'est pas nécessaire de très bien estimer les
probabilités \textit{a posteriori} pour bien prédire. En effet, prenons un
problème à deux classes, notées \(1\) et \(2\). Si
\(\P(Y=1|X=x) = 0.99\), l'estimateur de Bayes prédit alors la classe
\(1\) pour l'observation \(x\). Mais alors, si \(\hat{p}(x,1)=0.51\), le
prédicteur \(\hat{h}\) prédit lui aussi la classe \(1\) pour
l'observation \(x\), alors que l'estimation de la probabilité 
\textit{a posteriori}  est assez mauvaise.

\bigskip

Une des particularités des forêts aléatoires est qu'elles peuvent être
utilisées dans des cadres de régression et de classification, et seules
quelques légères adaptations sont nécessaires pour passer d'un cadre à
l'autre. De plus, elles présentent de très bonnes performances en
prédiction (c'est-à-dire en terme d'erreur de généralisation) dans les
deux cas.

\subsubsection{Une remarque sur l'espace d'entrée}

Souvent, on se limite à l'étude de problème
d'apprentissage statistique lorsque l'espace d'entrée \(\mathcal{X}\)
est égal à \(\mathbb{R}^p\), en particulier lorsqu'il est question de l'étude 
théorique des arbres ou des forêts aléatoires, accessible uniquement dans ce cas. 

L'entier naturel \(p\) désigne le nombre de
coordonnées de \(X\), et nous appelons ces coordonnées les variables 
explicatives ou les covariables du modèle et nous noterons \(X^j\) la \(j\)-ième variable.

Le rapport entre le nombre d'observations \(n\) et le nombre de
variables \(p\) est crucial en statistique, et peut mener à des
problèmes très différents (voir paragraphe \ref{forets-aleatoires-et-selection-de-variables}). 

\subsection{Un peu d'histoire}\label{un-peu-dhistoire}

Cette section présente quelques méthodes d'ensemble 
apparues avant les forêts aléatoires (détaillées plus bas) et souvent définies pour des règles 
de base qui ne sont pas des arbres de décision.

\subsubsection{Principe des méthodes d'ensemble}\label{Meth_Ens}

Les forêts aléatoires combinent des prédicteurs ou estimateurs de base que sont 
 les arbres,
donnant lieu à ce que l'on appelle aujourd'hui
les méthodes d'arbres. Plus largement ce sont des méthodes d'ensemble dont
le principe général (voir \cite{Dietter00}) est de construire une collection de 
prédicteurs, pour
ensuite agréger l'ensemble de leurs prédictions. En
régression, agréger les prédictions de \(q\) prédicteurs revient par
exemple à en faire la moyenne : chaque prédicteur fournit un
\(\hat{y}_l\), et la prédiction finale est alors
\(\frac{1}{q} \sum_{l=1}^q \hat{y}_l\). En classification,
l'agrégation consiste par exemple à faire un vote majoritaire parmi les labels 
des
classes fournis par les prédicteurs. 

Soulignons le fait que l'étape
d'agrégation de ces méthodes est toujours très simple et n'est pas
optimisée, contrairement aux méthodes dites d'agrégation de modèles,
qui pour une famille de prédicteurs donnée, cherche la meilleure manière
de les combiner pour obtenir un bon prédicteur agrégé (voir
par exemple les travaux de \cite{Lecue07}).

Ainsi, au lieu d'essayer d'optimiser une méthode ``en un coup'', les méthodes
d'ensemble génèrent plusieurs règles de prédiction et mettent ensuite en
commun leurs différentes réponses. L'heuristique de ces méthodes est
qu'en générant beaucoup de prédicteurs, on explore largement l'espace
des solutions, et qu'en agrégeant toutes les prédictions, on dispose d'un
prédicteur qui rend compte de cette exploration. Ainsi, on
s'attend à ce que le prédicteur final soit meilleur que chacun des
prédicteurs individuels. 

\noindent
Illustrons sur un cas simple en nous plaçant dans un cadre de classification à deux
classes. Pour que le classifieur agrégé commette une erreur pour un
\(x\) donné, il faut qu'au moins la moitié des classifieurs individuels
se soient également trompés pour ce même \(x\). On peut espérer que ceci
n'arrive pas très souvent, car même si les classifieurs individuels
commettent des erreurs, il est peu probable qu'ils commettent les mêmes
erreurs pour les mêmes entrées. D'où l'idée que les prédicteurs
individuels doivent être différents les uns des autres : la majorité ne
doit pas se tromper pour un  \(x\) donné. Pour que cela soit possible, il
faut également que les prédicteurs individuels soient relativement bons.
Et là où un prédicteur se trompe, les autres doivent prendre le relais
en ne se trompant pas.

\noindent
L'explication heuristique du succès de ces méthodes d'ensemble s'appuie donc sur 
deux propriétés : en premier lieu, chaque prédicteur individuel doit 
être relativement bon et 
en outre, les prédicteurs individuels doivent être différents les uns des 
autres.
Le premier point est nécessaire, car agréger des prédicteurs tous
mauvais ne pourra vraisemblablement pas donner un bon. Le
second point est également naturel, car agréger des prédicteurs trop peu variés 
donnera encore un prédicteur semblable et n'améliorera
pas les prédictions individuelles.

\subsubsection{Bagging}

Le Bagging est une méthode introduite par \cite{Breiman96} pour les arbres, et 
directement issue de la remarque 
selon laquelle les arbres CART sont instables et sensibles aux fluctuations de 
l'ensemble
des données de l'échantillon d'apprentissage \(\mathcal{L}_n\). On peut en
fait considérer plus généralement une méthode de prédiction (appelée
règle de base), qui construit sur \(\mathcal{L}_n\) un prédicteur
\(\hat{h}(.,\mathcal{L}_n)\). Le principe du Bagging est de tirer un
grand nombre d'échantillons, indépendamment les uns des autres, et de
construire, en appliquant à chacun d'eux la règle de base, un grand nombre de 
prédicteurs.
La collection de prédicteurs est alors agrégée en faisant simplement 
une moyenne ou un vote majoritaire. Cette méthode est centrale et on y reviendra 
largement dans la section \ref{bagging-1}.

\subsubsection{Boosting}

Introduit par \cite{Freund96}, le Boosting est une des méthodes
d'ensemble les plus performantes à ce jour. Étant donné un échantillon
d'apprentissage \(\mathcal{L}_n\) et une méthode de prédiction (ou
règle de base), qui construit sur \(\mathcal{L}_n\) un prédicteur
\(\hat{h}(.,\mathcal{L}_n)\). Le principe du Boosting est de tirer un
premier échantillon bootstrap \(\mathcal{L}_n^{\Theta_1}\), où chaque
observation a une probabilité \(1/n\) d'être tirée, puis d'appliquer la
règle de base pour obtenir un premier prédicteur
\(\hat{h}(.,\mathcal{L}_n^{\Theta_1})\). Ensuite, l'erreur de
\(\hat{h}(.,\mathcal{L}_n^{\Theta_1})\) sur l'échantillon
d'apprentissage \(\mathcal{L}_n\) est calculée. Un deuxième échantillon
boostrap \(\mathcal{L}_n^{\Theta_{2}}\) est alors tiré mais la loi du
tirage des observations n'est maintenant plus uniforme. La probabilité
pour une observation d'être tirée dépend de la prédiction de
\(\hat{h}(.,\mathcal{L}_n^{\Theta_1})\) sur cette observation. Le
principe est, par le biais d'une mise à jour exponentielle bien choisie,
d'augmenter la probabilité de tirer une observation mal
prédite et de diminuer celle de tirer une observation bien prédite. Une
fois le nouvel échantillon \(\mathcal{L}_n^{\Theta_{2}}\) obtenu, on
applique à nouveau la règle de base
\(\hat{h}(.,\mathcal{L}_n^{\Theta_{2}})\). On tire alors un troisième
échantillon \(\mathcal{L}_n^{\Theta_{3}}\), qui dépend des prédictions
de \(\hat{h}(.,\mathcal{L}_n^{\Theta_{2}})\) sur \(\mathcal{L}_n\) et
ainsi de suite. La collection de prédicteurs obtenus est alors agrégée
en faisant une moyenne pondérée, là encore via des poids exponentiels bien 
choisis.

Le Boosting est donc une méthode séquentielle, chaque échantillon étant
tiré en fonction des performances de la règle de base appliquée sur l'échantillon
précédent. En cela, le Boosting diffère de façon importante du Bagging,
où les échantillons sont tirés indépendamment les uns des autres, et
peuvent être obtenus en parallèle. L'idée du Boosting est de se concentrer de plus en plus sur les
observations mal prédites par la règle de base, pour essayer d'apprendre
au mieux cette partie difficile de l'échantillon en vue d'améliorer les
performances globales.

Mentionnons que le Boosting,  d’abord défini pour la classification, a été généralisé pour la 
régression par \cite{drucker1997improving}. On trouve dans ce cadre, une étude de son instabilité  dans 
\cite{gey2006boosting}  ou une utilisation pour la détection de données aberrantes dans 
 \cite{cheze2006iterated}.

Pour dénommer les méthodes de type Boosting, \cite{Breiman98} parle
d'algorithmes \textbf{Arc}ing, pour  \textbf{A}daptively  \textbf{R}esample and  \textbf{C}ombine.
L'idée est bien qu'au lieu de ré-échan- tillonner de façon indépendante
comme dans le Bagging, on ré-échantillonne de façon adaptative dans le
Boosting. Contrairement à nombre d'autres méthodes d'ensemble, le Boosting a
beaucoup été étudié théoriquement, voir par exemple \cite{Bartlett07}
et les références de cet article.

\subsubsection{Randomizing Outputs}

Le Bagging et le Boosting construisent une collection de prédicteurs en
ré-échantillonnant les observations de \(\mathcal{L}_n\), \cite{Breiman00} introduit la
méthode Randomizing Outputs pour les problèmes de régression, 
qui est une méthode d'ensemble de nature
différente. Le principe est, ici, de construire des échantillons
indépendants dans lesquels on altère les sorties de l'échantillon
d'apprentissage. La modification que subissent les sorties est obtenue
en rajoutant une variable de bruit à chaque \(Y_i\) de
\(\mathcal{L}_n\). On obtient alors une collection d'échantillons ``à
sorties randomisées'', puis on applique une règle de base sur chacun et
on agrège enfin l'ensemble des prédicteurs obtenus.

L'idée de Randomizing Outputs est, encore, qu'en appliquant une règle de
base sur des échantillons à sorties randomisées, on obtient une
collection de prédicteurs différents les uns des autres.

\subsubsection{Random Subspace}

 Un autre type de méthode d'ensemble est introduit dans \cite{Ho98}. Il n'est 
plus ici question de jouer sur l'échantillon, mais plutôt d'agir sur l'ensemble des
variables considérées. Le principe de la méthode Random Subspace est de
tirer aléatoirement un sous-ensemble de variables et d'appliquer une
règle de base sur \(\mathcal{L}_n\) qui ne prend en compte que les
variables sélectionnées. On génère alors une collection de prédicteurs
chacun construit en utilisant des variables différentes, puis on agrège
ces prédicteurs. Les sous-ensembles de variables sont tirés
indépendamment pour chaque prédicteur.

L'idée de cette méthode est de construire plusieurs prédicteurs, chacun
étant spécialisé et réputé bon dans un sous-espace de \(\mathcal{X}\) particulier, pour
ensuite en déduire un prédicteur sur l'espace d'entrée tout entier.

\subsubsection{Les ingrédients clés}

Les  méthodes d'ensemble évoquées ici ont toutes un principe général
commun. Il s'agit de partir d'une règle de prédiction de base, puis de
perturber cette règle, pour construire une collection de
prédicteurs issus de différentes perturbations, que
l'on agrège. Les perturbations
peuvent porter sur l'échantillon (ré-échantillonnage, sorties
randomisées) ou le sous-espace d'entrée dans lequel on construit le
prédicteur, et les différentes perturbations sont générées
indépendamment, ou non, les unes des autres.

Pour chacune de ces méthodes, les auteurs montrent, souvent sur des simulations,
que le prédicteur agrégé final fait systématiquement mieux (en terme
d'erreur de généralisation) que la règle de prédiction de base. Donc, en
pratique, il apparaît que ``perturber puis agréger'' améliore les
performances d'une méthode de prédiction donnée.

La remarque précédente ne vaut que si dans la collection construite, les
prédicteurs sont différents les uns des autres. C'est pourquoi ces
stratégies sont appliquées à des règles de base qui sont des méthodes dites ``instables''. Une
méthode est instable si de petites perturbations de l'échantillon
d'apprentissage peuvent engendrer de grandes modifications du prédicteur
obtenu. Par exemple, les arbres de décision sont instables au contraire
des méthodes linéaires (régression ou analyse discriminante) qui sont stables et donc jamais utilisées
comme règle de base dans les méthodes d'ensemble.

\subsection{Un jeu de données fil
rouge}\label{un-jeu-de-donnees-fil-rouge}

Tout au long de ce document, nous illustrerons l'application des différentes
méthodes sur les très classiques données  \ctt{spam}  à des fins pédagogiques.

Ce jeu de données, bien connu et largement 
disponible, est dû à un ingénieur de HP, prénommé George,  qui a analysé un 
échantillon de ses emails professionnels :

\begin{itemize}
\item Les individus sont les $4601$ emails en question dont $2788$ sont des emails souhaitables 
et $1813$ (soit 40\%) des emails
indésirables, c'est-à-dire des spams.
\item La variable réponse est donc binaire : 
\textit{spam} ou  \textit{non-spam}.
\item les variables explicatives sont au nombre de $p= 57$ : 
$54$ sont des proportions d'occurrences de mots ou de
  caractères, comme par exemple \ctt{\$}, \ctt{!}, \ctt{free}, \ctt{money}, $2$ sont liées 
  aux longueurs des suites de lettres majuscules (la moyenne, la plus longue) et 
  enfin la dernière est le nombre de lettres majuscules dans le mail. 
  Ces variables sont classiques et définies grâce à des procédures usuelles en analyse textuelle, 
  permettant de traiter statistiquement des individus caractérisés par des textes.
\end{itemize}
  
  Le but est double. Premièrement, construire un ``bon'' filtre anti-spam : un
  nouveau mail arrive, il faut réussir à prédire si c'est un spam ou
  non. Deuxièmement, on s'intéresse aussi à savoir quelles sont les
  variables sur lesquelles se base le plus le filtre anti-spam (ici
  des mots ou des caractères).
  
  Pour juger de la performance d'un filtre anti-spam, on
  découpe le jeu de données en deux : $2300$ mails pour
  l'apprentissage, $2301$ mails pour tester les prédictions.

Nous avons donc un problème de classification à 2 classes (\(L=2\)) avec
un nombre d'individus (\(n=2300\) pour l'apprentissage, la construction des modèles) largement plus
important que le nombre de variables (\(p=57\)). De plus, nous disposons
d'un échantillon test de grande taille (\(m=2301\)) pour évaluer une estimation de l'erreur de prédiction.

\section{Arbres CART}\label{arbres-cart}

L'acronyme CART signifie \textbf{C}lassification \textbf{A}nd \textbf{R}egression \textbf{T}rees. Il désigne
une méthode statistique, introduite par \cite{Breiman84} qui construit
des prédicteurs par arbre aussi bien en régression qu'en classification.

Parfois introduites avant CART, d'autres méthodes pour construire des arbres de 
décision sont connues et largement disponibles,
comme par exemple CHAID introduit par \cite{Kass1980} ou encore
l'algorithme \(C4.5\) par \cite{Quilan93}. On peut saisir 
encore aujourd'hui l'actualité des arbres de décision dans deux récentes références :
\cite{Patil2012} en informatique et \cite{Loh2014} en statistique, qui 
contiennent des synthèses utiles et les bibliographies associées.

Dans la suite, bien qu'en principe plusieurs façons de construire des  
arbres de décision CART sont possibles, par exemple en
changeant la famille de coupures autorisées, la fonction de coût ou
encore la règle d'arrêt, nous nous limitons à celle, couramment
utilisée, présentée dans le livre de \cite{Breiman84}. Nous y renvoyons
pour des compléments et des variantes qui n'ont pas vraiment eu la diffusion
de la version la plus simple dont  l'interprétabilité, en particulier, a fait 
le succès. Un exposé en français, concis et 
clair de la méthode CART en régression peut être trouvé dans le chapitre 2
de la thèse de \cite{Gey02}.

Le principe général de CART est de partitionner récursivement l'espace
d'entrée \(\mathcal{X}\) de façon binaire, puis de déterminer une
sous-partition optimale pour la prédiction. Bâtir un arbre CART se fait en deux étapes.
Une première phase est la construction d'un arbre maximal, qui  permet de définir la famille de 
modèles à l'intérieur de laquelle on cherchera à sélectionner le meilleur, et une
seconde phase, dite d'élagage, qui construit une suite de sous-arbres optimaux
élagués de l'arbre maximal. Détaillons chacune de ces étapes.

\subsection{Construction de l'arbre
maximal}\label{construction-de-larbre-maximal}

\begin{figure}
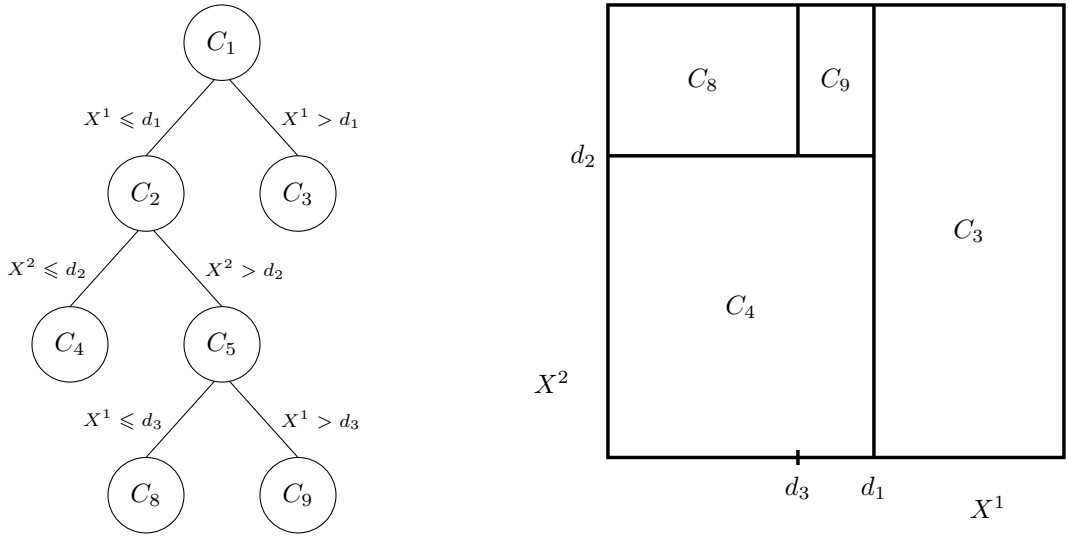

\begin{minipage}[ht]{0.5\textwidth}
  \begin{pgfpicture}{0 cm}{0 cm}{6 cm}{7 cm}
    \pgfcircle[stroke]{\pgfxy(2,0.5)}{0.5 cm}
    \pgfcircle[stroke]{\pgfxy(4,0.5)}{0.5 cm}
    \pgfcircle[stroke]{\pgfxy(1,2.5)}{0.5 cm}
    \pgfcircle[stroke]{\pgfxy(3,2.5)}{0.5 cm}
    \pgfcircle[stroke]{\pgfxy(2,4.5)}{0.5 cm}
    \pgfcircle[stroke]{\pgfxy(4,4.5)}{0.5 cm}
    \pgfcircle[stroke]{\pgfxy(3,6.5)}{0.5 cm}
    
    \pgfputat{\pgfxy(2,0.5)}{\pgfbox[center,center]{$C_8$}}
    \pgfputat{\pgfxy(4,0.5)}{\pgfbox[center,center]{$C_9$}}
    \pgfputat{\pgfxy(1,2.5)}{\pgfbox[center,center]{$C_4$}}
    \pgfputat{\pgfxy(3,2.5)}{\pgfbox[center,center]{$C_5$}}
    \pgfputat{\pgfxy(2,4.5)}{\pgfbox[center,center]{$C_2$}}
    \pgfputat{\pgfxy(4,4.5)}{\pgfbox[center,center]{$C_3$}}
    \pgfputat{\pgfxy(3,6.5)}{\pgfbox[center,center]{$C_1$}}
    
    \pgfputat{\pgfxy(1.7,1.5)}{\pgfbox[center,center]{\scriptsize $X^1 \leq 
d_3$}}
    \pgfputat{\pgfxy(4.3,1.5)}{\pgfbox[center,center]{\scriptsize $X^1 > d_3$}}
    \pgfputat{\pgfxy(0.7,3.5)}{\pgfbox[center,center]{\scriptsize $X^2 \leq 
d_2$}}
    \pgfputat{\pgfxy(3.3,3.5)}{\pgfbox[center,center]{\scriptsize $X^2 > d_2$}}
    \pgfputat{\pgfxy(1.7,5.5)}{\pgfbox[center,center]{\scriptsize $X^1 \leq 
d_1$}}
    \pgfputat{\pgfxy(4.3,5.5)}{\pgfbox[center,center]{\scriptsize $X^1 > d_1$}}
    
    \pgfxyline(2.9,6.01)(2,5)
    \pgfxyline(3.1,6.01)(4,5)
    \pgfxyline(1.9,4.01)(1,3)
    \pgfxyline(2.1,4.01)(3,3)
    \pgfxyline(2.9,2.01)(2,1)
    \pgfxyline(3.1,2.01)(4,1)
  \end{pgfpicture}
\end{minipage}
\begin{minipage}[ht]{0.5\textwidth}
    \begin{pgfpicture}{-1 cm}{-1 cm}{6 cm}{6 cm}
        \pgfsetlinewidth{1.3 pt}
        \pgfrect[stroke]{\pgfxy(0,0)}{\pgfxy(6,6)}
        \pgfxyline(2.5,4)(2.5,6)
        \pgfxyline(3.5,0)(3.5,6)
        \pgfxyline(0,4)(3.5,4)
        \pgfxyline(2.5,0.1)(2.5,-0.1)
        
        \pgfputat{\pgfxy(2.5,-0.5)}{\pgfbox[center,base]{$d_3$}}
        \pgfputat{\pgfxy(3.5,-0.5)}{\pgfbox[center,base]{$d_1$}}
        \pgfputat{\pgfxy(-0.15,4)}{\pgfbox[right,center]{$d_2$}}
        
        \pgfputat{\pgfxy(5,-0.8)}{\pgfbox[center,base]{$X^1$}}
        \pgfputat{\pgfxy(-0.5,1)}{\pgfbox[right,center]{$X^2$}}
        
        \pgfputat{\pgfxy(1.75,2)}{\pgfbox[center,center]{$C_4$}}
        \pgfputat{\pgfxy(4.75,3)}{\pgfbox[center,center]{$C_3$}}
        \pgfputat{\pgfxy(1.25,5)}{\pgfbox[center,center]{$C_8$}}
        \pgfputat{\pgfxy(3,5)}{\pgfbox[center,center]{$C_9$}}
    \end{pgfpicture}
\end{minipage}
\caption{\`A gauche : un arbre de classification permettant de prédire le label
correspondant à un $x$ donné. \`A droite : la partition associée dans l'espace des variables explicatives }\label{fig::tree}
\end{figure}

A chaque pas du partitionnement, on découpe une partie de l'espace en
deux sous-parties. On associe alors naturellement un arbre binaire à la
partition construite. Les n\oe{}uds de l'arbre sont associés aux éléments
de la partition. Par exemple, la racine de l'arbre est associé à
l'espace d'entrée tout entier. Ses deux n\oe{}uds fils sont associés aux
deux sous-parties obtenues par la première découpe du partitionnement,
et ainsi de suite. La figure \ref{fig::tree} illustre la
correspondance entre un arbre binaire et la partition associée.

Détaillons maintenant la règle de découpe. Pour fixer les idées, le lecteur peut se restreindre
à des variables explicatives continues (la cas qualitatif est néanmoins mentionné explicitement
dans le texte chaque fois que cela est utile), l'espace d'entrée est alors \(\mathbb{R}^p\), où
\(p\) est le nombre de variables. Partons de la racine de l'arbre
(associée à \(\mathbb{R}^p\) tout entier), qui contient toutes les
observations de l'échantillon d'apprentissage \(\mathcal{L}_n\). La
première étape de CART consiste à découper au mieux cette racine en deux
n\oe{}uds fils. Nous appelons coupure (ou découpe ou même split) un élément de la 
forme

\[\{ X^j \leq d \} \cup \{ X^j > d \},\]

\noindent où \(j \in \{1,\ldots,p\}\) et \(d\in \mathbb{R}\). Découper
suivant \(\{ X^j \leq d \} \cup \{ X^j > d \}\) signifie que toutes les
observations avec une valeur de la \(j\)-ième variable plus petite que
\(d\) vont dans le n\oe{}ud fils de gauche, et toutes celles avec une
valeur plus grande que \(d\) vont dans le n\oe{}ud fils de droite. La
méthode sélectionne alors la meilleure découpe, c'est-à-dire le couple
\((j,d)\) qui minimise une certaine fonction de coût :

\begin{itemize}
    \item En régression, on cherche à minimiser la variance
    intra-groupes résultant de la découpe d'un n\oe{}ud $t$ en 2 n\oe{}uds fils $t_L$ et $t_R$.
    La variance d'un n\oe{}ud $t$ étant définie par $V(t)=\frac{1}{\#t}\sum_{i : x_i \in t} (y_i - \overline{y}_t)^2$ où $\overline{y}_t$ est la moyenne des $y_i$ des 
observations présentes dans le n\oe{}ud $t$ et l'on est donc conduit à minimiser
$$\frac{1}{n}\sum_{(x_i,y_i) \in t_L}(y_i-\overline{y}_{t_L})^2+\frac{1}{n}\sum_{(x_i,y_i) \in t_R}(y_i-\overline{y}_{t_R})^2 =\frac{\#t_L}{n}V(t_L)+\frac{\#t_R}{n}V(t_R) .$$
 
   \item En classification (où l'ensemble des classes est $\{1,\ldots,L\}$), on 
définit l'impureté des n\oe{}uds fils, le plus souvent par le biais de l'indice de Gini. L'indice de Gini d'un 
n\oe{}ud $t$ est défini par $\Phi(t)=\sum_{c=1}^L \hat{p}^c_t (1 - \hat{p}^c_t)$, où 
$\hat{p}^c_t$ est la proportion d'observations de classe $c$ dans le n\oe{}ud $t$.
On est alors conduit à pour tout n\oe{}ud $t$ et tout split admissible à maximiser 
 $$\Phi(t)-\left( \frac{\#t_L}{\#t}\Phi(t_L)+\frac{\#t_R}{\#t}\Phi(t_R)\right).$$

\end{itemize}

\noindent En régression, on cherche donc des découpes qui tendent à
diminuer la variance des n\oe{}uds obtenus. En classification, on cherche à
diminuer la fonction de pureté de Gini, et donc à augmenter l'homogénéité des n\oe{}uds
obtenus, un n\oe{}ud étant parfaitement homogène s'il ne contient que des
observations de la même classe. Au passage, on pourrait imaginer de mesurer 
l'homogénéité des n\oe{}uds par une autre fonction liée au taux de 
fausses classifications mais ce
choix naturel ne conduit pas à une fonction de pureté strictement 
concave garantissant l'unicité de l'optimum à chaque découpe.

Mentionnons ici que dans le cas d'une variable explicative $X^j$ catégorielle, 
rien ce qui précède ne change sauf que dans ce cas, une coupure est simplement un élément de la forme 
\[\{ X^j \in d \} \cup \{ X^j \in \bar{d} \},\] où $d$ et $\bar{d}$ sont non vides et constituent une partition de l'ensemble
des modalités de la variable $X^j$.

Une fois la racine de l'arbre découpée, on se restreint à chacun des
n\oe{}uds fils et on recherche alors, suivant le même procédé, la meilleure
façon de les découper en deux nouveaux n\oe{}uds, et ainsi de suite. Les
arbres sont ainsi développés, jusqu'à atteindre une condition d'arrêt. Une
règle d'arrêt classique consiste à ne pas découper des n\oe{}uds qui
contiennent moins d'un certain nombre d'observations. Les n\oe{}uds
terminaux, qui ne sont plus découpés, sont appelés les feuilles de
l'arbre. A noter, que l'on ne découpe pas un n\oe{}ud pur, c'est-à-dire un
n\oe{}ud ne contenant que des observations dont les sorties sont les mêmes
(typiquement en classification). On appelle arbre maximal, noté $T_{max}$, l'arbre
pleinement développé. Dans le même temps, on associe à chaque n\oe{}ud
\(t\) de l'arbre une valeur (\(\overline{Y}_t\) en régression ou le label de la classe
majoritaire des observations présentes dans le n\oe{}ud \(t\) en
classification). Donc, à un arbre est associée une partition (définie
par ses feuilles) et également des valeurs qui sont attachées à chaque élément de
cette partition. Le prédicteur par arbre est alors la fonction constante
par morceaux, associée à l'arbre (voir figure \ref{fig::tree}).

\subsection{Élagage}\label{elagage}

La deuxième étape de l'algorithme CART, s'appelle l'élagage et consiste
à chercher le meilleur sous-arbre élagué de l'arbre maximal (meilleur au
sens de l'erreur de généralisation). L'idée est que l'arbre maximal
possède une très grande variance et un biais faible. \textit{A contrario}, un
arbre constitué uniquement de la racine (qui engendre alors un
prédicteur constant) a une très petite variance mais un biais élevé.
L'élagage est une procédure de sélection de modèles, où les modèles sont
les sous-arbres élagués de l'arbre maximal, soit tous les sous-arbres binaires de
$T_{max}$ ayant la m\^eme racine que $T_{max}$. Cette procédure minimise un critère
pénalisé où la pénalité est proportionnelle au nombre de feuilles de
l'arbre. 

Tous les sous-arbres binaires de $T_{max}$ 
contenant la racine sont des mod\`{e}les admissibles. Entre
 $T_{max}$, le mod\`{e}le de complexit\'{e} maximale, qui conduit au surajustement 
aux donn\'{e}es de l'\'{e}chantillon d'apprentissage et l'arbre restreint à la racine qui est fortement biaisé,
il s'agit de trouver l'arbre optimal parmi les admissibles. 
En nombre fini, il suffirait donc au moins en principe, de construire 
la suite de tous les meilleurs arbres 
à $k$ feuilles pour $1 \leq k \leq |T_{max}|$, où  $|T|$ désigne le nombre de feuilles de l'arbre $T$, et de les comparer par exemple sur 
un échantillon test. Mais le nombre de modèles admissibles est exponentiel d'où une 
complexit\'{e} algorithmique explosive. Fort heureusement, une énumération implicite et efficace suffit pour atteindre 
un résultat 
optimal. Le moyen consiste simplement dans l'algorithme d'\'{e}lagage qui
assure l'extraction d'une 
suite de sous-arbres embo\^{i}t\'{e}s  (c'est-à-dire élagués les uns des autres) $T_1,\ldots,T_K$ tous \'{e}lagu\'{e}s de 
$T_{max}$, o\`{u} 
$T_k$ minimise un crit\`{e}re des moindres carr\'{e}s p\'{e}nalis\'{e} en 
régression. Cette suite est obtenue de mani\`ere it\'erative en coupant des branches à chaque \'etape, ce 
qui ram\`ene la complexit\'e \`a un niveau tr\`es raisonnable. On se restreint sans 
inconvénient au cas de la régression
dans les quelques lignes qui suivent, la situation étant identique en 
classification. 

La clé est de pénaliser l'erreur d'ajustement d'un sous-arbre $T$ \'{e}lagu\'{e} 
de $T_{max}$ : $$\overline{err}(T) = \frac{1}{n}\sum_{\{t \ feuille \ de \ T\}} 
\sum_{(x_i,y_i) \in t}(y_i-\overline{y}_t)^2$$
par une fonction linéaire du nombre de feuilles $|T|$ conduisant au 
critère des moindres carrés pénalisés : $$crit_{\alpha}(T) = 
\overline{err}(T)+\alpha |T|.$$

\noindent
Ainsi $\overline{err}(T)$ qui mesure l'{ajustement} du 
mod\`{e}le $T$ aux donn\'{e}es, d\'{e}cro\^{i}t avec le nombre de feuilles alors 
que $|T|$ qui quantifie la {complexit\'{e}} du mod\`{e}le $T$, cro\^{i}t avec le 
nombre de feuilles. Le paramètre $\alpha$ r\'{e}gle  la p\'{e}nalit\'{e} : plus 
$\alpha$ est grand, plus les mod\`{e}les complexes c'est-à-dire comptant beaucoup de feuilles, sont 
p\'{e}nalis\'{e}s.

L'algorithme d'\'{e}lagage est donné dans la Table \ref{Algo_elagage}, où pour 
tout n\oe{}ud interne $t$ d'un arbre $T$, on note $T_t$ la branche de $T$ issue 
du n\oe{}ud $t$ (contenant tous les descendants du n\oe{}ud $t$) et l'erreur correspondante
est donnée par $\overline{err}(t)=n^{-1}\sum_{\{x_i\in t\}}(y_i-\bar{y}_t)^2$.

\begin{table}[ht!]
\begin{center}
\begin{tabular}{| r | l |}
\hline
 {\bf Entr\'{e}e} & Arbre maximal $T_{max}$.\\
                 & \\
{\bf Initialisation} & $\alpha_1=0$, $T_1=T_{\alpha_1}=\argmin_{T \ 
\mbox{\'{e}lagu\'{e} de} \ T_{max}}\overline{err}(T)$.\\
                           & {\tt initialiser} $T=T_1$ et $k=1$.\\
                           & \\
{\bf Iteration}   & {\tt Tant que $|T|>1$}, \\
                       & \hspace{0.5cm} {\tt Calculer}\\
                       & \hspace{0.8cm} $\alpha_{k+1} = \displaystyle{\min_{\{t 
\ \mbox{n\oe{}ud interne de} \ 
T\}}\frac{\overline{err}(t)-\overline{err}(T_t)}{|T_t|-1}}$.\\
                       & \hspace{0.5cm} {\tt Elaguer} toutes les branches $T_t$ 
de $T$ telles que \\
                       & \hspace{0.8cm} 
$\overline{err}(T_t)+\alpha_{k+1}|T_t|=\overline{err}(t)+\alpha_{k+1}$\\
                       & \hspace{0.5cm} {\tt Prendre} $T_{k+1}$ le sous-arbre 
\'{e}lagu\'{e} ainsi obtenu.\\
                       & \hspace{0.5cm} {\tt Boucler sur}  $T=T_{k+1}$ et $k = 
k+1$.\\
                       & \\
{\bf Sortie}      & Arbres $T_1\succ \ldots  \succ T_K=\{t_1\}$,\\
                        & Param\`{e}tres $(0=\alpha_1;\ldots; \alpha_K)$.\\
\hline
\end{tabular}
\caption{Algorithme d'élagage de CART}
\label{Algo_elagage}
\end{center}
\end{table}

Le résultat principal du livre \cite{Breiman84} établit que la suite de 
param\`{e}tres $(0 = \alpha_1; \ldots; \alpha_K)$ est strictement croissante, associée à la
suite $T_1\succ \ldots  \succ T_K=\{t_1\}$  constituée de modèles emboités 
au sens de l'élagage et que, pour tout $1\leq k\leq K$ 
\begin{eqnarray*}
\forall \alpha \in [\alpha_{k}, \alpha_{k+1}[  \quad \quad \ T_k & = & \argmin_{\{T \ 
\mbox{sous-arbre de} \ T_{max}\}}crit_{\alpha}(T)\\
         & = & \argmin_{\{T \ \mbox{sous-arbre de} \ 
T_{max}\}}crit_{\alpha_k}(T)
\end{eqnarray*}
en posant ici $\alpha_{K+1}=+\infty$.

Autrement dit, la suite $T_1, \ldots, T_K$ contient  toute l'information 
statistique utile puisque pour tout $\alpha\geqslant 0$, le sous-arbre 
minimisant $crit_{\alpha}$ est un sous-arbre de la suite produite par 
l'algorithme d'élagage.

On peut la visualiser (voir l'exemple sur les données \ctt{spam} dans la Figure  
\ref{suite_optimale}) par le biais de la suite des param\`{e}tres 
$(\alpha_k)_{1\leq k\leq K}$, avec le nombre de feuilles, les erreurs de 
validation (obtenues par validation croisée) de chaque arbre, et une estimation de l'\'{e}cart-type 
de cette erreur. Chaque point repr\'{e}sente ainsi un arbre, avec l'estimation 
de l'\'{e}cart-type de 
l'erreur de validation sous forme de segment vertical. Le choix de l'arbre 
optimal peut se faire 
directement en minimisant l'erreur de validation ou en appliquant la r\`{e}gle du 1 
écart-type consistant à choisir
l'arbre le plus compact atteignant une erreur de validation inférieure à la valeur la valeur du 
minimum précédent 
augmenté de l'écart-type estimé de l'erreur de validation. Cette quantité est représentée par la ligne horizontale en 
pointillés sur l'exemple de la Figure \ref{suite_optimale}.

Il faut remarquer que, bien entendu, si un arbre quelconque de cette suite comporte $k$ 
feuilles, c'est
le meilleur arbre à $k$ feuilles. En revanche, cette suite ne contient pas tous les 
meilleurs arbres 
à $k$ feuilles pour $1 \leq k \leq |T_{max}|$ mais seulement une partie d'entre 
eux. Les arbres manquants ne sont simplement  pas compétitifs car de critère pénalisé plus 
grand.

Comme nous le verrons plus bas, les forêts aléatoires sont, la plupart du temps, 
des forêts d'arbres non
élagués. Cependant,  insistons sur le fait
qu'un arbre CART, s'il est utilisé seul, doit être élagué.

\subsection{Garantie théorique}\label{garantie-theorique}

Une première garantie théorique minimale est disponible dans \cite{Breiman84},
puisqu'en annexe figure un résultat assez général de consistance d'arbres de décision mais dont les 
conditions ne sont pas nécessairement vérifiées, en particulier par l'arbre 
élagué optimal. Il faut attendre \cite{gey2005model}  pour obtenir en régression un résultat non asymptotique 
justifiant la forme de la pénalité et \cite{gey2012risk}  pour un
résultat semblable dans le cas de la classification, et c'est celui-ci que nous 
 proposons d'esquisser ci-dessous. 

Il s'agit des bornes de risque sur l'étape d'élagage pour la classification binaire. 
Le critère pénalisé est alors de la forme :
$$\hat{R}_{pen}(T)=\frac{1}{n}\sum_{i=1}^n\mathbf{1}_{\hat{f}_T(X_i)\neq
Y_i}+\alpha |T|$$
avec, comme ci-dessus, $|T|$ le nombre de feuilles de $T$.
Un résultat typique pour la classification binaire dit que lorsque 
le sous-arbre $T_{opt}$ est choisi par la méthode du {Hold-out} avec un 
échantillon $\mathcal{L}_1$  
pour construire et élaguer $T_{max}$, un échantillon $\mathcal{L}_2$ pour 
choisir l'arbre minimisant l'erreur de  prédiction et
sous une condition sur la marge $h$, il existe des
constantes positives $C_1,C_2,C_3$ telles que :
 $$\E\left[l(f^*,\hat{f}_{T_{opt}}) | \mathcal{L}_1 \right] \leq C_1
\inf_{T\preceq T_{max}} \left[ \inf_{f\in S_T} l(f^*,f) +
h^{-1} \frac{|{T}|}{n_1} \right] + \frac{C_2}{n_1} + C_3
\frac{\ln n_1}{n_2}$$
 où $S_T$ est l'ensemble des classifieurs définis sur la 
 partition $\Tf$ induite par l'ensemble des feuilles de  $T$, 
 et $l(f^*,f) = \P\left(f(X) \neq Y\right)-\P\left(f^*(X)\neq 
Y\right)$.

Ce résultat établit que la performance effective de l'arbre sélectionné est, au 
premier ordre,
du même ordre de grandeur que la performance du meilleur classifieur augmentée 
de la pénalité, en justifiant ainsi la forme. 

Un point  à remarquer est que la 
qualité de la sélection de l'estimateur est appréciée conditionnellement à 
l'échantillon $\mathcal{L}_1$, 
puisque la famille de modèles
à l'intérieur de laquelle on fouille est dépendante des données. Le risque 
de référence est évalué sur la famille $S_T$ est l'ensemble des classifieurs définis sur la  
 partition $\Tf$, issus de l'arbre maximal. 
 
 \subsection{Interprétabilité et
instabilité}\label{interpretabilite-et-instabilite}

On rassemble dans cette section des remarques dont l'utilité est manifeste pour 
le statisticien appliqué.

\subsubsection{Découpes compétitives}

On peut disposer en chaque n\oe{}ud de l'arbre de la suite ordonnée par réduction 
décroissante de l'hétérogénéité de toutes les découpes (une par variable explicative).
C'est ce que l'on appelle les découpes compétitives ou concurrentes (competing splits) 
et elles sont, en tout n\oe{}ud, nécessairement calculées lors de la construction de l'arbre maximal.  
La possibilité du développement manuel de l'arbre maximal peut être précieuse et s'opère 
en choisissant en chacun des n\oe{}uds dans la liste ordonnée 
des splits, soit la découpe optimale, soit une découpe légèrement moins bonne
mais portant sur des variables moins incertaines, plus faciles, moins chères à 
mesurer ou encore plus interprétables (voir par exemple \cite{ghattas2000agregation}).

\subsubsection{Valeurs manquantes}

L'une des difficultés concrètes pour le calcul d'une prédiction est la présence de
valeurs manquantes. CART en offre un traitement efficace et très élégant. En effet, tout d'abord
remarquons que lorsque des variables sont manquantes pour un $x$ donné, 
cela ne pose un problème que si l'on passe par un n\oe{}ud dont la coupure est basée 
sur l'une de ces variables. Ensuite, en un n\oe{}ud ou la variable sur laquelle porte le split
est manquante, on peut penser utiliser l'une des autres variables, par
exemple la seconde découpe compétitive. Mais cette idée n'est pas optimale, 
puisque 
la règle d'acheminement dans les fils droit et gauche respectivement peut
être très éloignée de la règle d'acheminement induite par la découpe optimale. 
D'où l'idée de calculer en chaque n\oe{}ud la liste des découpes de
substitution (surrogate splits) définies par la découpe minimisant le nombre de 
désaccords avec la règle d'acheminement induite par la coupure optimale. On 
dispose ainsi d'une méthode de traitement des valeurs manquantes en prédiction
locale et  performante, qui ne passe pas par des méthodes d'imputation globales et
souvent trop grossières.

 Notons que si CART permet de bien gérer les valeurs manquantes  en prédiction, 
les forêts aléatoires, qui sont des ensembles d'arbres non élagués, perdent 
essentiellement cette propriété. En effet, tout d'abord, pour un  arbre non élagué donné, le chemin de la racine 
à une feuille est en général beaucoup plus long et, de plus,  potentiellement de nombreux arbres sont 
touchés par une variable manquante particulière.
Des solutions existent néanmoins dans ce cadre avec les forêts 
aléatoires 
floues comme par exemple dans  \cite{Valdiviezo2015}. Bien sûr, mais c'est une autre question, on peut 
utiliser les RF pour faire de l'imputation comme dans \cite{stekhoven2012missforest}.

\subsubsection{Interprétabilité}

L'interprétabilité est un des aspects qui font le succès des arbres CART. On peut, en effet, 
très facilement répondre à la 
question du pourquoi, pour un $x$ donné, on prévoit telle valeur de $y$ en fournissant
la suite des réponses aux questions constituées par les découpes successives 
rencontrés pour parcourir, pour le $x$ en question, l'unique chemin de la racine à la 
feuille associée. Notons que les forêts aléatoires perdent cette propriété d'interprétabilité puisqu'il
n'existe pas d'arbre associé à une forêt, même si l'on peut signaler des tentatives pour
définir un arbre approchant en un certain sens une forêt,
comme par exemple \cite{hara2016making}. 

Mais de manière plus globale, au delà de l'interprétation d'une prédiction particulière, on peut,
une fois l'arbre CART construit, considérer que les variables qui interviennent 
dans les découpes des n\oe{}uds de l'arbre (en particulier les n\oe{}uds
les plus proches de la racine) sont les variables les plus utiles pour
le problème considéré. En effet, plus une découpe est proche de la racine plus la 
décroissance de l'hétérogénéité est importante. 
C'est une façon heuristique et pratique d'avoir une
information sur l'importance des variables. En réalité, cette première
intuition donne des résultats biaisés et un indice d'importance des
variables plus élaboré est fourni par les arbres CART. Il est basé à nouveau sur 
la notion de découpes de substitution. Suivant \cite{Breiman84}, on peut définir
l'importance d'une variable en évaluant, en chaque n\oe{}ud, la réduction 
d'hétérogénéité engendrée par l'usage du split de substitution portant sur cette variable 
puis en les sommant sur tous les n\oe{}uds. Cet indice est, par exemple, utilisé par \cite{Poggi06} dans une procédure 
de sélection de variables, laquelle est examinée théoriquement par \cite{sauve2014variable}. Pour intéressant qu'il soit, cet indice d'importance 
des variables n'est plus aujourd'hui très utilisé : il est en effet peu 
intuitif, très instable car dépendant d'un arbre donné et moins pertinent que l'importance
des variables par permutation au sens des forêts aléatoires. En outre, son analogue,
qui n'utilise pas les découpes de substitution, existe dans 
les forêts mais il a tendance à favoriser les variables catégorielles qui ont un grand 
nombre de modalités (nous y reviendrons en \ref{variantes-et-extensions-1}).

\subsubsection{Valeurs aberrantes}

Un autre avantage est la résistance naturelle aux valeurs aberrantes, la 
méthode 
étant purement non paramétrique, la présence d'une donnée aberrante dans 
l'ensemble d'apprentissage  va contaminer essentiellement la feuille qui la 
contient, 
avec un faible impact pour les autres.

A propos des données aberrantes ou plus largement des données influentes, on 
peut utiliser l'instabilité de CART pour mesurer l'influence, non pas des 
variables, 
mais des individus sur l'analyse, cette approche classique en analyse des 
données
est porteuse d'informations sur les individus qui ne sont pas seulement vus comme des répétitions uniquement utiles pour apprendre 
sur le modèle ou les variables. Ainsi dans \cite{BarHen2015} 
sont proposées diverses mesures de l'influence des observations sur les 
résultats
obtenus avec un arbre de classification CART. Ces mesures d'influence 
quantifient la sensibilité de
l'analyse pour un arbre de classification de référence construit avec toutes les données
par des mesures de la variabilité des 
prédictions fourni par les arbres jackknife construits avec toutes les observations moins 
une. 
L'analyse est étendue aux séquences d'arbres élagués pour produire une
notion d'influence spécifique à la méthode CART.

\subsubsection{Complexité algorithmique}

D'autres qualités peuvent encore être notées et sans  plus
développer   citons-en une dernière : le temps de calcul,
qui en fait une méthode bien adaptée à l'analyse de vastes ensembles de
données. En effet, l'algorithme de construction d'un arbre et l'algorithme 
d'élagage sont de 
complexité informatique faible : le nombre d'opérations requise est  
$\mathcal{O}(pn log(n))$ au mieux 
(arbre parfaitement équilibré) ou de $\mathcal{O}(pn^2$) au pire (arbre dit en peigne ou en arête 
de poisson, le plus déséquilibré).

\subsection{Exemple}\label{exemple}

Examinons cette construction sur le jeu de données \ctt{spam}. 
Le package utilisé ici pour réaliser les arbres CART est le package
\ctt{rpart} \citep{Therneau13} inclus dans la base de \ctt{R} \citep{RCT_R2015}.

\begin{figure}[!ht]
\begin{center}
\includegraphics[width=0.8\textwidth]{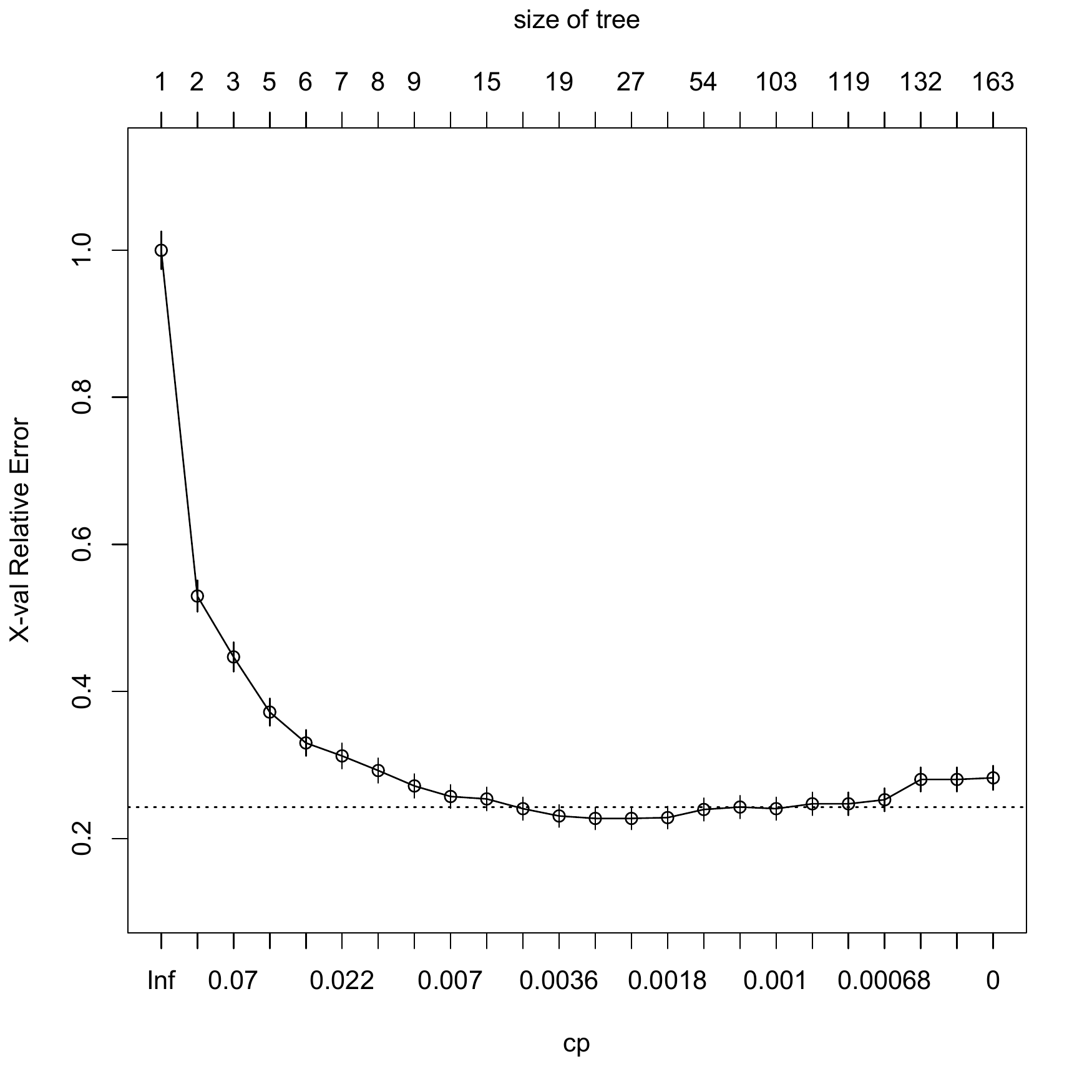}
  \caption{La suite optimale des sous-arbres élagués de $T_{max}$}
  \label{suite_optimale}
\end{center}
\end{figure}

La suite optimale des sous-arbres élagués $T_{max}$ obtenue par la 
construction de l'arbre maximal
et l'application de l'algorithme d'élagage est donné
par la Figure \ref{suite_optimale}. 

\begin{figure}[!ht]
\begin{center}
\hspace*{-10ex}
\includegraphics[width=1.15\textwidth]{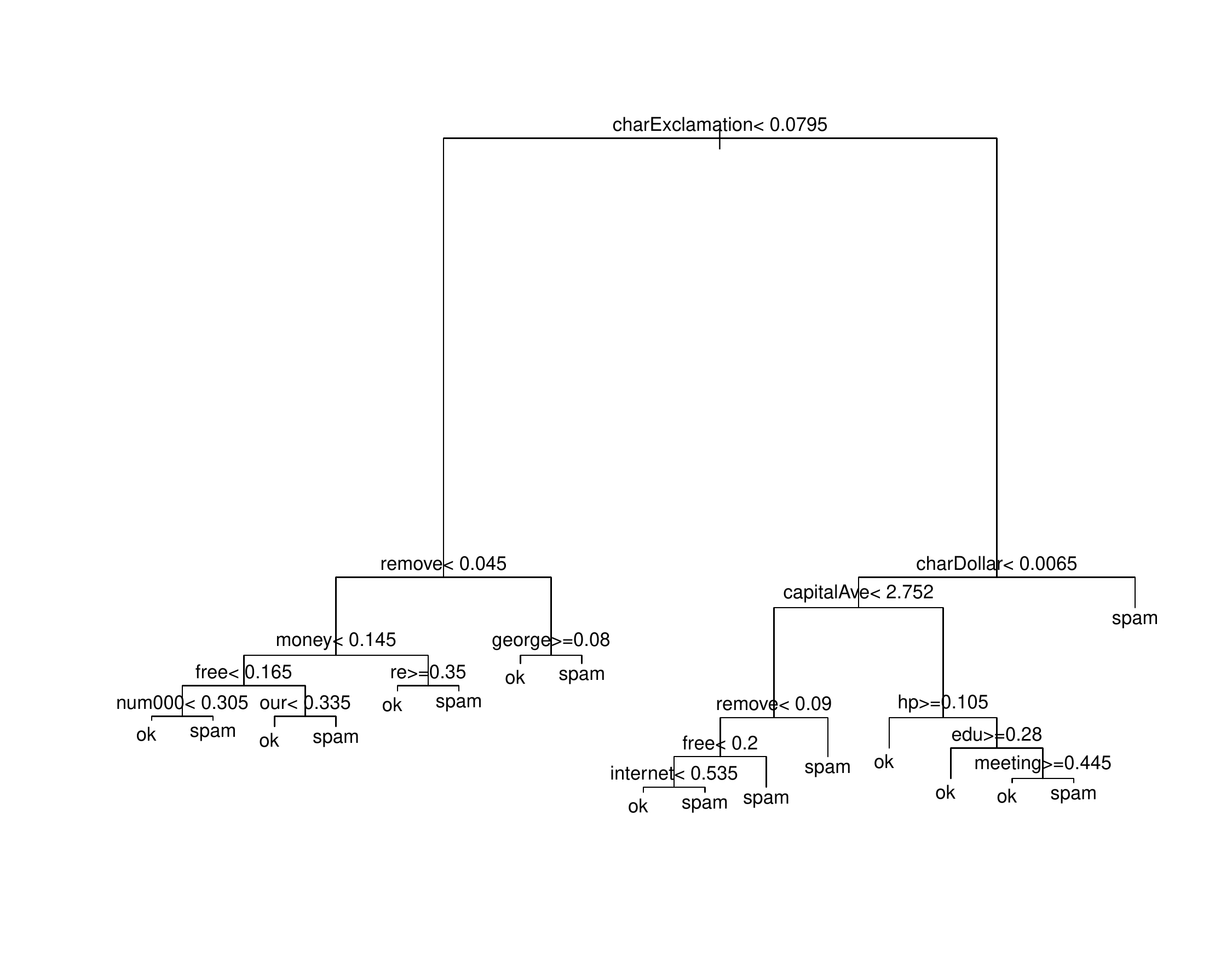}
  \vspace*{-10ex}\caption{Arbre CART élagué}
  \label{ArbreCARTelague}
\end{center}
\end{figure}

Chaque point repr\'{e}sente ainsi un arbre, avec l'estimation de 
l'\'{e}cart-type de 
l'erreur de validation croisée sous forme de segment vertical. 
L'arbre minimisant ce critère est parfois encore un peu trop complexe 
(\(23\)
feuilles ici), et en relaxant un peu la condition de minimisation de
l'erreur de prédiction par la règle du ``1 écart-type'' (1 s.e. rule de Breiman 
qui tient compte de l'incertitude de l'estimation de l'erreur des arbres de la suite), 
on obtient l'arbre de la Figure \ref{ArbreCARTelague}.

Le meilleur sous-arbre élagué de l'arbre maximal (à un
écart-type près)  comporte 17 feuilles et seules 14 variables,
parmi les 57 initiales, figurent dans les découpes associées 
aux 16 n\oe{}uds internes :
\texttt{charExclamation}, \texttt{charDollar}, \texttt{remove}, \texttt{capitalAve}, \texttt{money}, \texttt{george}, \texttt{hp}, 
\texttt{free}, \texttt{re}, \texttt{num000}, \texttt{our}, \texttt{edu}, \texttt{internet} et \texttt{meeting}.

On peut illustrer la facilité d'interprétation en considérant par exemple le  
chemin de la racine à 
la feuille la plus à droite qui dit qu'un mail qui contient beaucoup de \ctt{\$} et 
de \ctt{!} est presque 
toujours un spam. Inversement le  chemin de la racine à 
la cinquième feuille la plus à droite exprime qu'un mail contenant beaucoup 
de \ctt{!}, de lettres capitales et 
de \ctt{hp} mais peu de \ctt{\$}  n'est presque jamais un spam. Pour cette dernière interprétation, il 
n'est pas inutile de rappeler que les mails examinés sont les mails professionnels d'un seul individu 
travaillant pour HP. 

Enfin les erreurs empiriques et tests obtenues par différents arbres sont données par la Table 
\ref{perfos}.

\begin{table}[!ht]
  \centering
  \begin{tabular}{c|cccc}
    Arbre & 2 feuilles & 1 s.e. & maximal & optimal \\
    \hline
    Erreur empirique & 0.208 & 0.073 &
    0.000 & 0.062 \\
    Erreur test & 0.209 & 0.096 & 0.096 & 0.086
  \end{tabular}
  \caption{Erreurs empirique et test des 4 arbres pour les données 
\ctt{spam}}
  \label{perfos}
\end{table}

On remarque que, comme annoncé, l'arbre maximal (trop complexe) a une erreur 
empirique (\textit{i.e.} sur l'échantillon 
d'apprentissage) nulle et que l'arbre à deux feuilles (trop simple) présente des erreurs 
test et empirique proches. L'arbre optimal,
quant à lui, a la meilleure erreur test de $0.086$, soit $8,6\%$.

\subsection{Extensions}\label{extensions}

De nombreuses extensions ou variantes des arbres CART ont été proposées pour diverses
fins.

On trouve tout d'abord des extensions, en régression, qui 
construisent des prédicteurs plus réguliers que les prédicteurs par arbres qui sont constants par
morceaux, par exemple, l'algorithme MARS introduit par \cite{Friedman91}. 

D'autres idées pour déterminer les splits, en les choisissant
déterministes, ce qui évite la dépendance de la famille de modèles aux données
évoquée plus haut. Dans  \cite{Donoho1997}, pour des applications 
en traitement d'images, les découpes considérées sont dyadiques. 
Le découpage dyadique d'un rectangle du plan est déterministe et complet
jusqu'à la résolution du pixel, c'est l'analogue de l'arbre maximal et 
l'élagage est opéré grâce à un algorithme classique 
de choix de la meilleure base de paquets d'ondelettes. Des idées semblables ont
été généralisées dans \cite{Blanchard2007}. 

Citons aussi l'une des extensions les plus utilisées : les méthodes CART pour les données de survie par 
exemple \cite{LeBlanc1993} et \cite{Molinaro2004} ainsi que le plus récent article de synthèse  
\cite{BouHamad2011}. 

Une extension aux données spatiales est à mentionner dans \cite{bel2009cart} avec des 
applications dans le domaine de l'environnement.

On peut enfin renvoyer au livre récent \cite{zhang2013recursive} portant plus 
largement sur les 
méthodes basées sur le partitionnement récursif, mais qui présente aussi des variantes 
de CART pour les données
longitudinales ou pour les données fonctionnelles. Dans cette ligne, signalons
l'utilisation de CART en chimiométrie dans \cite{questier2005use}.

\section{Des arbres aux forêts
aléatoires}\label{des-arbres-aux-forets-aleatoires}

\subsection{Définition générale des forêts
aléatoires}

Après avoir examiné le socle des méthodes d'arbres, abordons le coeur de ce papier. Les forêts aléatoires ont été introduites par \cite{Breiman01} par la 
définition très générale suivante :

\begin{definition}\label{def_gen_rf}
  Soit $\left( \widehat{h}(., \Theta_1), \ldots, \widehat{h}(.,
    \Theta_q) \right)$ une collection de prédicteurs par arbres, avec
  $\Theta_1, \ldots, \Theta_q$ $q$ variables aléatoires
  i.i.d. indépendantes de $\mathcal{L}_n$. Le prédicteur des forêts aléatoires $\widehat{h}_{RF}$ est obtenu
  est agrégeant cette collection d'arbres aléatoires de la façon suivante :
  \begin{itemize}
  \item $\widehat{h}_{RF} (x) = \frac{1}{q} \sum_{l=1}^q
    \widehat{h}(x, \Theta_l)$ (moyenne des prédictions individuelles
    des arbres) en régression,
  \item $\widehat{h}_{RF} (x) = \amax{1\leq k \leq K} \sum_{l=1}^q
    \mathds{1}_{\widehat{h}(x, \Theta_l) = k}$ (vote majoritaire
    parmi les prédictions individuelles des arbres) en classification.
  \end{itemize}
\end{definition}

En fait, cette définition n'est pas rigoureusement identique à celle de
\cite{Breiman01}, qui ne précise pas que les variables $\Theta_l$ sont
indépendantes de $\mathcal{L}_n$. Nous adoptons néanmoins la
Définition~\ref{def_gen_rf} car elle reflète mieux l'intuition que l'aléa
supplémentaire apporté par les $\Theta_l$ est déconnecté des données
d'apprentissage et de plus toutes les variantes de forêts aléatoires
connues rentrent dans ce cadre.

\begin{figure}[!ht]
\begin{center}
  \includegraphics[width=0.7\textwidth]{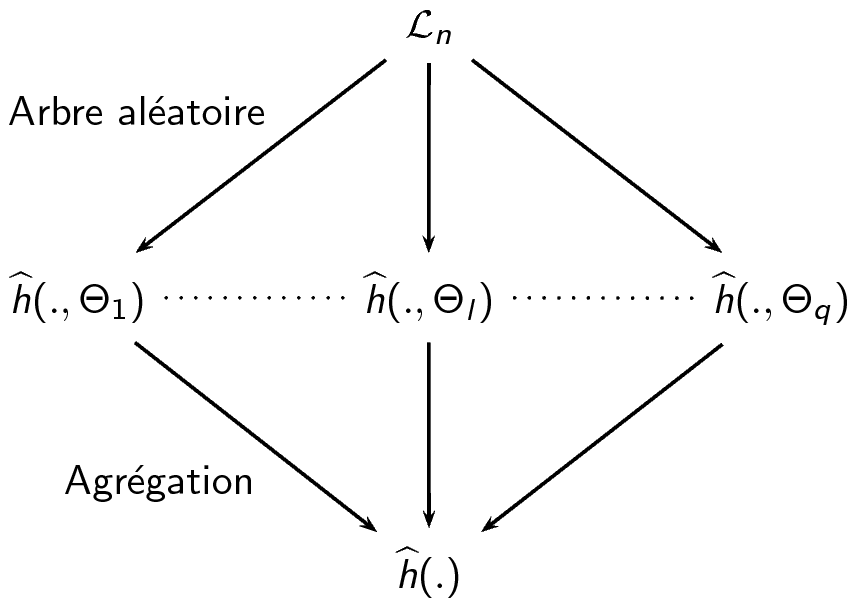}
   \caption{Schéma général des forêts aléatoires}
 \label{SchemaGeneralRF}
\end{center}
\end{figure}

Cette définition est illustrée par le schéma de la
Figure~\ref{SchemaGeneralRF}. Nous en verrons plusieurs déclinaisons par
la suite. 

Le terme forêt aléatoire vient du fait que les prédicteurs individuels
sont, ici, explicitement des prédicteurs par arbre, et du fait que
chaque arbre dépend d'une variable aléatoire supplémentaire
(c'est-à-dire en plus de \(\mathcal{L}_n\)). Une forêt aléatoire est
l'agrégation d'une collection d'arbres aléatoires.

Les forêts aléatoires font bien partie de la famille des méthodes
d'ensemble. Remarquons d'ailleurs que, parmi les méthodes d'ensemble
précédemment citées (lorsque l'on choisit comme règle de base un arbre de
décision), seul le Boosting ne rentre pas dans la définition de forêts
aléatoires. En effet, les arbres individuels du Boosting ne dépendent
pas d'aléas indépendants les uns des autres. Le Bagging, Randomizing
Outputs et Random Subspace sont alors des cas particuliers de forêts
aléatoires, avec respectivement pour aléa supplémentaire le tirage de
l'échantillon bootstrap, la modification aléatoire des sorties de
\(\mathcal{L}_n\) et le tirage des sous-ensembles de variables. En plus
de ces trois méthodes, il existe de nombreux cas particuliers de forêts
aléatoires dans la littérature. 

Il existe une ambiguïté
de vocabulaire dans la littérature. En effet, Leo Breiman, dans son article de 2001,
définit les forêts aléatoires comme ci-dessus et sont donc pour lui une famille de méthodes. Or, dans le même article, il
présente un cas particulier de forêts aléatoires, appelées Random
Forests-RI, qu'il a implémentées (voir \cite{BreiCutl04}). Par la suite, ce
sont ces Random Forests-RI qui ont été quasi-systématiquement utilisées
dans d'innombrables applications réelles. Et pour cause, le
programme est accessible à tous,  facile d'utilisation et la méthode
atteint des performances empiriques exceptionnelles. Finalement, la dénomination
``forêts aléatoires'' désigne maintenant très souvent les Random
Forests-RI. On trouve également le terme de ``forêts aléatoires de Leo
Breiman'' pour désigner les Random Forests-RI.

Dans la suite de cette section, nous nous concentrons tout d'abord sur le Bagging 
qui est central dans l'analyse en évoquant quelques résultats avant de décrire  
en détails les Random Forests-RI et de définir l'erreur OOB qui en est
une sortie très utile. Nous donnons ensuite quelques idées des garanties 
théoriques disponibles pour des variantes plus ou moins éloignées des RF-RI mais 
utiles. Enfin nous terminerons comme d'habitude par l'application aux données \ctt{spam}
avant d'évoquer variantes et extensions. 

\subsection{Bagging}\label{bagging-1}

La méthode du Bagging a été introduite par \cite{Breiman96}. Le mot
Bagging est la contraction des mots \textbf{B}ootstrap et
\textbf{Agg}regat\textbf{ing}. Étant donné un échantillon
d'apprentissage \(\mathcal{L}_n\) et une méthode de prédiction (appelée
règle de base), qui construit sur \(\mathcal{L}_n\) un prédicteur
\(\hat{h}(.,\mathcal{L}_n)\). Le principe du Bagging est de tirer
indépendamment plusieurs échantillons bootstrap
\(( \mathcal{L}_n^{\Theta_1}, \ldots, \mathcal{L}_n^{\Theta_q} )\),
d'appliquer la règle de base sur chacun d'eux pour obtenir une
collection de prédicteurs
\((\hat{h}(.,\mathcal{L}_n^{\Theta_1}), \ldots, 
\hat{h}(.,\mathcal{L}_n^{\Theta_q}))\),
et enfin d'agréger ces prédicteurs de base.

L'idée du Bagging, et qu'en appliquant la règle de base sur différents
échantillons bootstrap, on en modifie les prédictions, et donc on
construit ainsi une collection variée de prédicteurs. L'étape
d'agrégation permet alors d'obtenir un prédicteur performant.

\begin{figure}[!ht]
\begin{center}
  \includegraphics[width=0.7\textwidth]{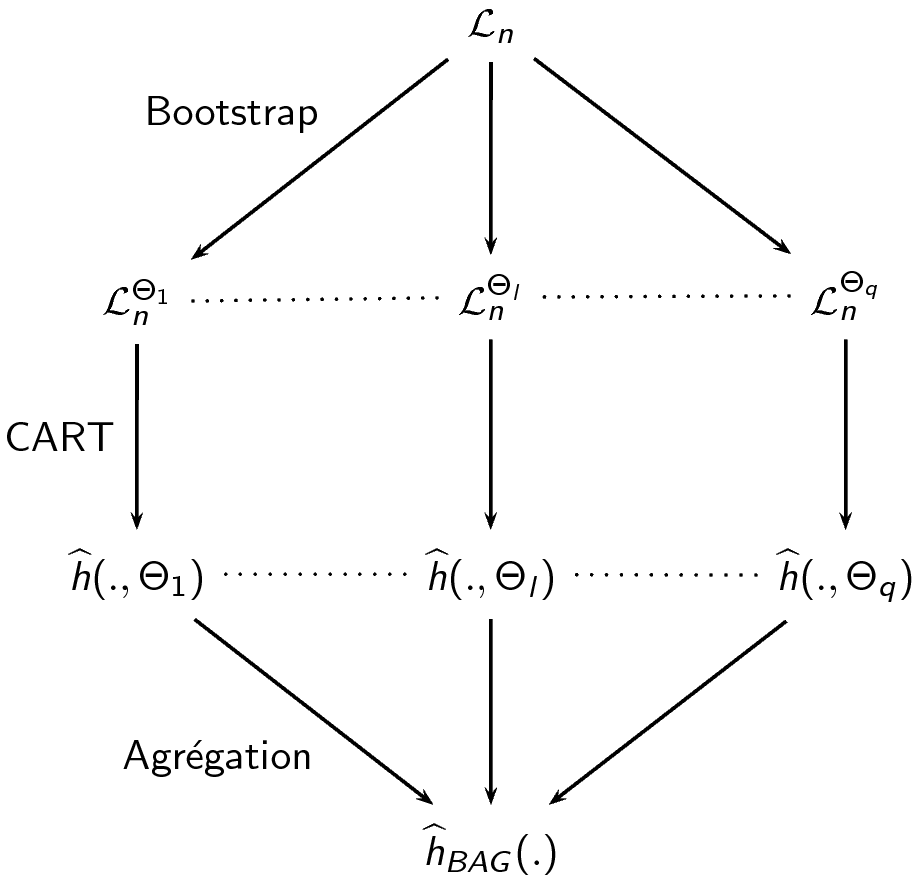}
  \caption{Schéma du Bagging avec pour règle de base un arbre CART}
  \label{SchemaBagging}
\end{center}
\end{figure}

Un échantillon bootstrap \(\mathcal{L}_n^{\Theta_l}\) est, par exemple,
obtenu en tirant aléatoirement \(n\) observations avec remise dans
l'échantillon d'apprentissage \(\mathcal{L}_n\), chaque observation
ayant une probabilité \(1/n\) d'être tirée. La variable aléatoire
\(\Theta_l\) représente alors ce tirage aléatoire. Une deuxième façon
classique d'obtenir un échantillon bootstrap est de tirer aléatoirement
\(k\) observations sans remise dans \(\mathcal{L}_n\), avec \(k<n\).
La Figure \ref{SchemaBagging} résume le principe de cette méthode.

Initialement, le Bagging a été introduit avec comme règle de base un
arbre de décision. Cependant, le
schéma est très général et peut-être appliqué à d'autres règles de base
comme par exemple, la règle du plus proche voisin. Cette méthode du plus
proche voisin ``baggé'' a été étudiée, dans un cadre de
régression, par \cite{Biau10b}, puis \cite{Biau10a} (voir également les
références de cet article). Le premier article établit la consistance de
la méthode du plus proche voisin ``baggé'' (l'estimateur obtenu converge
vers la vraie fonction de régression quand \(n\) tend vers \(+\infty\)),
à condition que le nombre d'observations \(k\) dans les échantillons
bootstrap (avec ou sans remise) tende vers \(+\infty\), mais moins vite
que \(n\) : \(\frac{k}{n} \to +\infty\). Ces conditions de convergence
sur \(k\) et \(n\) sont des conditions que nous retrouverons plus bas. 
Le deuxième article va plus loin
et montre que l'estimateur atteint la vitesse optimale de convergence
pour la classe des fonctions Lipschitz, sous les mêmes conditions sur
\(k\) et \(n\).

Cette étude illustre les bienfaits des méthodes d'ensemble :
partant d'une règle de base assez pauvre (la règle du plus proche voisin
n'est pas consistante), le Bagging la transforme en une règle aux très
bonnes propriétés asymptotiques (consistance et vitesse optimale de
convergence). L'idée, ici, est que la méthode du plus proche voisin,
n'explore pas assez l'espace : elle assigne à un \(x\) donné, le \(y\)
correspondant à l'observation de \(\mathcal{L}_n\) la plus proche de
\(x\). Le fait d'appliquer la méthode, sur un échantillon bootstrap permet
de prendre en compte les sorties des observations plus éloignées de
\(x\) (ce qui arrive lorsque les plus proches voisins de \(x\) ne sont
pas présents dans l'échantillon bootstrap courant). Le plus proche
voisin ``baggé'' met alors un poids sur chacune des données de
\(\mathcal{L}_n\) et le prédicteur agrégé est finalement une moyenne
pondérée des \(Y_i\) de l'échantillon d'apprentissage. Les résultats
théoriques nous assurent en fait que la méthode règle automatiquement et
de façon optimale ces poids.

Nous évoquons maintenant un article sur l'analyse du Bagging.
\cite{Buhlmann02} étudient, en dimension \(1\)
(\(\mathcal{X} = \mathbb{R}\)), la convergence des trois paramètres qui
définissent un arbre à deux feuilles (le point de coupure et les deux
valeurs assignées à chacune des feuilles). 
Ces trois paramètres convergent en loi à la vitesse \(n^{1/3}\)
\citep{Banerjee07} et \cite{Buhlmann02} obtiennent alors un résultat
montrant que le prédicteur Bagging d'arbres à deux feuilles
(avec un sous-échantillonnage sans remise à la place du bootstrap) a une 
variance inférieure à celle de l'arbre à deux feuilles initial.
C'est ainsi le premier résultat théorique illustrant l'intuition que les
méthodes d'ensemble permettent une réduction de la variance.

\subsection{Forêts aléatoires ``Random
Inputs''}\label{forets-aleatoires-random-inputs}

\begin{figure}[!ht]
\begin{center}
  \includegraphics[width=0.8\textwidth]{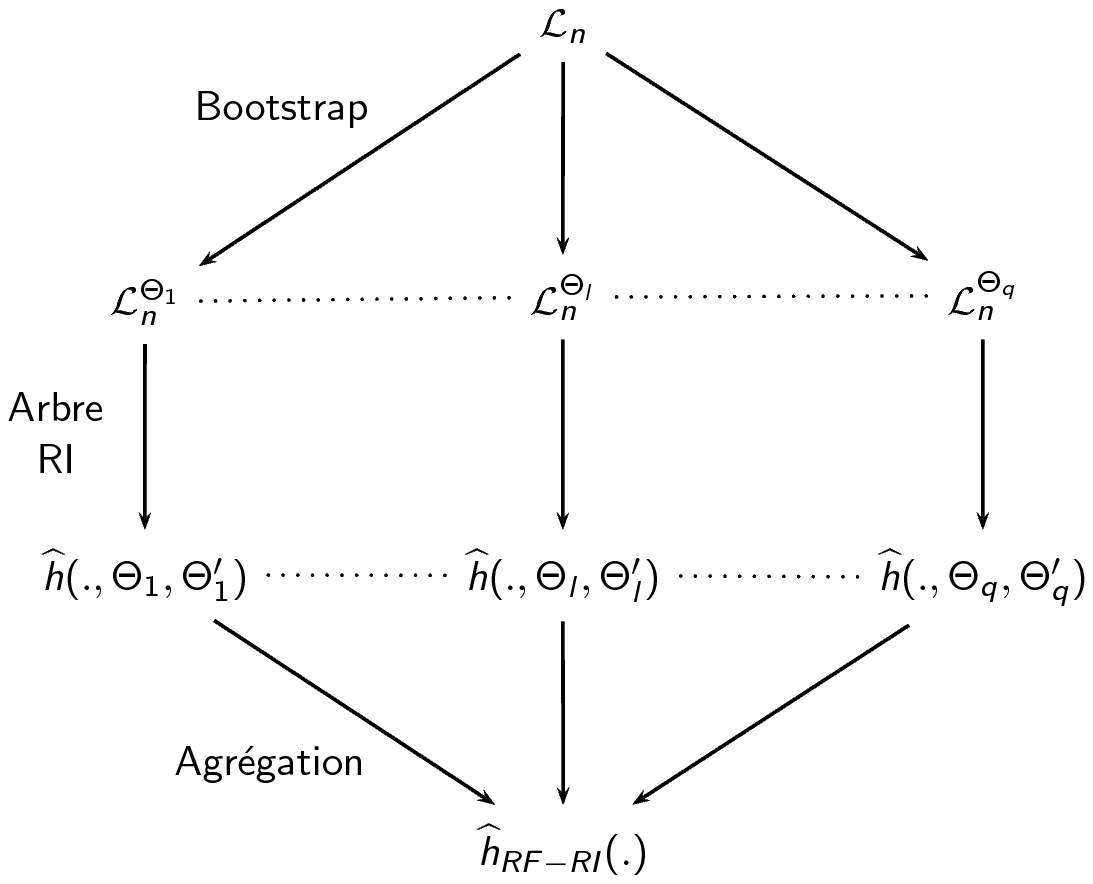}
  \caption{Schéma des forêts aléatoires RF-RI}
  \label{SchemaRFRI}
\end{center}
\end{figure}

Random Forests-RI signifie ``forêts aléatoires à variables d'entrée
aléatoires'' (Random Forests with Random Inputs) et le principe de
leur construction est tout d'abord de générer plusieurs échantillons bootstrap
\(\mathcal{L}_n^{\Theta_1}, \ldots, \mathcal{L}_n^{\Theta_q}\) (comme
dans le Bagging). Ensuite, sur chaque échantillon
\(\mathcal{L}_n^{\Theta_l}\), une variante de CART est appliquée. Plus
précisément, un arbre est, ici, construit de la façon suivante. Pour
découper un n\oe{}ud, on tire aléatoirement un nombre \(m\) de variables,
et on cherche la meilleure coupure uniquement suivant les \(m\)
variables sélectionnées. De plus, l'arbre construit est complètement
développé (arbre maximal) et n'est pas élagué. La collection d'arbres
obtenus est enfin agrégée (moyenne en régression, vote majoritaire en
classification) pour donner le prédicteur Random Forests-RI.
La Figure \ref{SchemaRFRI} fournit
le schéma récapitulatif de l'algorithme RF-RI, où \(\Theta\)
désigne le tirage bootstrap et \(\Theta'\) désigne le tirage aléatoire
des variables.

Ainsi, les Random Forests-RI peuvent être vues comme une variante du
Bagging, où la différence intervient dans la construction des arbres
individuels (les étapes de bootstrap et d'agrégation étant les mêmes).
Le tirage, à chaque n\oe{}ud, des \(m\) variables se fait, sans remise, et
uniformément parmi toutes les variables (chaque variable a un
probabilité \(1/p\) d'être choisie). Le nombre \(m\) (\(m \leq p\)) est
fixé au début de la construction de la forêt et est donc identique pour
tous les arbres  et pour tous les n\oe{}uds d'un même arbre
mais naturellement les $m$ variables impliquées dans 
deux n\oe{}uds distincts sont en général différentes. C'est un paramètre très important de la méthode. Une
forêt construite avec \(m=p\) revient à faire du Bagging d'arbres CART
non élagués, alors qu'une forêt construite avec \(m=1\) est très
différente du Bagging. En effet, lorsque \(m=1\), le choix de la
variable, suivant laquelle est découpé un n\oe{}ud, est complètement
aléatoire (les coupures suivant cette variable ne sont pas mises en
compétition avec des coupures utilisant d'autres variables).

Le tirage des \(m\) variables à chaque n\oe{}ud représente un aléa
supplémentaire, par rapport au Bagging. Pour les Random Forests-RI, il y
a donc deux sources d'aléas pour générer la collection des prédicteurs
individuels : l'aléa dû au bootstrap et l'aléa du choix des variables
pour découper chaque n\oe{}ud d'un arbre. Ainsi, on perturbe à la fois
l'échantillon sur lequel on lance la règle de base et le
coeur de la construction de la règle de base. Ce tirage aléatoire de
variables pour découper un n\oe{}ud avait déjà été utilisé par
\cite{Amit97} dans des problèmes de reconnaissance d'image. Leur méthode
a beaucoup influencé Leo Breiman dans sa mise au point de Random
Forests-RI. Pour leur problème, le nombre de coupures candidates était
tellement gigantesque qu'ils étaient obligés de réduire le nombre de
possibilités, par exemple en effectuant un choix aléatoire préliminaire
à la découpe.

En pratique, les Random Forests-RI (avec le paramètre \(m\) bien choisi)
améliorent les performances du
Bagging (voir la comparaison des méthodes sur des données de référence
dans \cite{Breiman01}). L'explication heuristique de ces améliorations
est que le fait de rajouter un aléa supplémentaire pour construire les
arbres, rend ces derniers encore plus différents les uns des autres,
sans pour autant dégrader de façon significative leurs performances
individuelles. Le prédicteur agrégé est alors meilleur. Nous avons vu
que la plupart des méthodes d'ensemble construisent une collection de
prédicteurs qui sont des versions perturbées d'une règle de base. La
perturbation introduite doit alors réaliser un compromis : une trop grande perturbation dégrade les prédicteurs
individuels et le prédicteur agrégé est alors mauvais alors qu'une trop petite
perturbation induit des prédicteurs individuels trop similaires entre
eux et le prédicteur agrégé n'apporte alors aucune amélioration. Les
excellents résultats des Random Forests-RI en pratique laissent penser
qu'elles (avec le paramètre \(m\) bien choisi) réalisent un bon
compromis, en injectant la ``bonne dose'' d'aléa.

L'algorithme des Random Forests-RI a été codé par \cite{BreiCutl04}. Il a
ensuite été importé dans le logiciel libre \texttt{R} 
par \cite{Liaw02}, via le
package \texttt{randomForest}. Ce package est
utilisé dans le traitement de très nombreuses applications réelles. 
Il existe deux principaux paramètres dans ce programme et qui sont 
aussi les deux seuls véritables paramètres de la méthode :

\begin{itemize}
    \item Le paramètre le plus important est le nombre $m$ de variables choisies 
aléatoirement à chacun des n\oe{}uds des arbres. Il est nommé \texttt{mtry} dans le 
package. Il peut varier de $1$ à $p$ et possède une valeur par défaut : 
$\sqrt{p}$ en classification, $p/3$ en régression. 

A cet égard on trouve dans \cite{Gen08}), 
une étude empirique qui précise les valeurs par défaut : pour les problèmes de 
régression, à l'exception du temps de calcul, il n'y a pas d'amélioration par rapport à Bagging 
non élagué (obtenu pour $mtry = p$). Pour les problèmes de classification standards, la valeur 
par défaut proposée dans le package est bon mais pour des problèmes de classification de 
grande dimension, des valeurs plus grandes pour $mtry$ donnent parfois des résultats bien meilleurs.
   \item Nous pouvons également jouer sur le nombre d'arbres $q$ de la forêt. Ce 
paramètre est nommé \texttt{ntree} et sa valeur par défaut est $500$.
\end{itemize}

Le programme permet également de régler d'autres aspects de la méthode :
le nombre minimum d'observations (nommé \texttt{nodesize}) en dessous
duquel on ne découpe pas un n\oe{}ud, ou encore la façon d'obtenir les
échantillons bootstrap (avec ou sans remise, ainsi que le nombre
d'observations tirées). Nous laisserons dans les expériences numériques,  les valeurs par
défaut pour ces éléments, \textit{i.e.}~un \texttt{nodesize} de \(1\) en classification et \(5\) en
régression, et les échantillons bootstrap considérés sont tous obtenus
en tirant \(n\) observations avec remise dans l'échantillon
d'apprentissage \(\mathcal{L}_n\).

\subsection{Erreur OOB}\label{erreur-oob}

En plus de construire un prédicteur, l'algorithme des Random Forests-RI
calcule une estimation de son erreur de généralisation : l'erreur
Out-Of-Bag (OOB) où ``Out-Of-Bag'' signifie  ``en dehors du
bootstrap''. Cette erreur était déjà calculée par l'algorithme du
Bagging, d'où la présence du mot ``Bag''. Le procédé de calcul de cette
erreur est le suivant et il est très astucieusement intriqué dans l'algorithme de 
construction d'une RF.

Fixons une observation \((X_i,Y_i)\) de l'échantillon d'apprentissage
\(\mathcal{L}_n\) et considérons l'ensemble des arbres
construits sur les échantillons bootstrap ne contenant pas cette
observation, c'est-à-dire pour lesquels cette observation est
``Out-Of-Bag''. Nous agrégeons alors uniquement les prédictions de ces
arbres pour fabriquer notre prédiction \(\hat{Y}_i\) de \(Y_i\). Après
avoir fait cette opération pour toutes les données de \(\mathcal{L}_n\),
nous calculons alors l'erreur commise : l'erreur
quadratique moyenne en régression
\(\left( \frac{1}{n} \sum_{i=1}^n ( \hat{Y}_i - Y_i )^2 \right)\), et la
proportion d'observations mal classées en classification
\(\left( \frac{1}{n} \sum_{i=1}^n \mathds{1}_{ \hat{Y}_i \neq Y_i } \right)\).
Cette quantité est appelée erreur OOB du prédicteur Random Forests-RI.

Cette estimation obéit aux mêmes contraintes que celles des
estimateurs classiques de l'erreur de généralisation (par échantillon test ou
par validation croisée), au sens où les données prédites sont des données
qui n'ont pas été rencontrées au préalable par le prédicteur utilisé. Un
avantage de l'erreur OOB par rapport aux estimateurs classiques est
qu'elle ne nécessite pas de découpage de l'échantillon d'apprentissage.
Ce découpage est en quelque sorte inclus dans la génération des
différents échantillons bootstrap. Cependant, il faut bien noter que
pour chaque observation ce n'est pas le même ensemble d'arbres qui est
agrégé. En conséquence, cette erreur estime l'erreur de généralisation
d'une forêt, mais elle n'utilise jamais les prédictions de la forêt
elle-même, mais plutôt celles de prédicteurs qui sont des agrégations
d'arbres de cette forêt. Par la suite, nous
adopterons ici quasi exclusivement l'erreur OOB puisqu'elle est très liée à la définition
de l'importance des variables et que nous l'utiliserons pour comparer des prédicteurs entre
eux, et non pour obtenir une estimation précise de leurs erreurs de
généralisation. Bien entendu, dans ce dernier cas, on procédera plus classiquement
grâce à un échantillon test.

\subsection{Garantie théorique}\label{garantie-theorique-1}

En dehors d'une majoration assez générale mais relativement grossière due à
\cite{Breiman01}, il n'y a pas de résultats théoriques disponibles pour les
RF-RI à proprement parler. C'est l'une des rares, mais elle est cruciale,
difficultés de cette méthode. On peut cependant citer des résultats 
théoriques, dus à un petit nombre d'auteurs, qui considèrent différentes
variantes des forêts aléatoires et fournissent un premier pas vers des 
garanties théoriques tout en esquissant une compréhension plus profonde des
extraordinaires performances pratiques des RF-RI, sans toutefois les élucider
pour le moment.

En confondant ces différentes variantes (que l'on examinera en partie en
\ref{variantes-et-extensions}), on peut trouver
des résultats de consistance dans \cite{breiman2000}, 
\cite{biau2008consistency},  \cite{zhu2015reinforcement},
\cite{ishwaran2010consistency} et \cite{denil2014narrowing} ;
un résultat de réduction de variance et de
vitesse de convergence en dimension 1 dans  \cite{genuer2012variance}, 
un résultat de réduction du biais et de vitesse de convergence en dimension
quelconque dans \cite{arlot2014analysis} ; et enfin un résultat de vitesse
de convergence dans un contexte de réduction de dimension dans \cite{biau_JMLR2012}.

Signalons une contribution récente de \cite{scornet2015consistency} portant
sur la consistance d'une variante assez réaliste des RF-RI dans le contexte
des modèles additifs. Enfin, on peut trouver des résultats de normalité
asymptotique dans \cite{mentch2014quantifying} ou encore dans 
\cite{wager2014asymptotic}, ainsi que des résultats de vitesse de convergence
pour des forêts en lien avec des estimateurs à noyaux dans
\cite{scornet2016random}.

Sur ces questions, on ne saurait trop recommander la lecture du récent papier
de revue \cite{biau2016random} qui est orienté vers les questions de l'analyse
théorique des RF, ainsi que la discussion qui suit, à laquelle l'un d'entre
nous a contribué (voir \cite{arlot2016comments}). Un objectif de ce commentaire
est de quantifier l'apport des différents ingrédients des forêts aléatoires,
tout d'abord de manière théorique pour une variante très simple de forêt, puis
par simulation sur une variante très proche des RF-RI. Il ressort de cette
étude que c'est la randomisation des partitions (qu'elle soit obtenue grâce au
bootstrap, au tirage des $m$ variables à chaque n\oe{}ud ou au tirage du point de
coupure) associées aux arbres de la forêt qui serait la plus cruciale.
Ceci explique (au moins partiellement) pourquoi les méthodes du Bagging (qui ne randomise pas sur la recherche de la coupure) et Extra-Trees
de \cite{Geurts06} (qui n'utilise pas de bootstrap, voir
\ref{variantes-et-extensions})
donnent toutes deux des résultats très satisfaisants en pratique
alors qu'elles sont très différentes dans le choix de l'aléa
supplémentaire $\Theta$.

\subsection{Exemple}\label{exemple-1}

Les taux d'erreur test des prédicteurs Bagging et forêts aléatoires sont
regroupés dans le tableau~\ref{perfos_bag_rf} et comparés à la performance
de l'arbre CART optimal.

\begin{table}[!ht]
  \centering
  \begin{tabular}{c|ccc}
    Prédicteur & arbre optimal & bagging & forêt aléatoire  \\
    \hline
    Erreur test & 0.086 & 0.060 & 0.052
  \end{tabular}
  \caption{Erreurs test du bagging et des forêts aléatoires, comparées à celles de l'arbre optimal pour les données \ctt{spam}}
  \label{perfos_bag_rf}
\end{table}

Pour appliquer le Bagging sur les données \ctt{spam}, nous utilisons le
package \ctt{randomForest} et construisons donc un prédicteur Bagging
avec comme règle de base un arbre CART non-élagué (le package ne permet pas
d'élaguer les arbres d'une forêt).

Le Bagging est réputé stabiliser l'arbre CART classique en
réduisant fortement sa variance et en améliorer sensiblement les performances. 
Sur notre jeu de données, l'erreur test atteinte par le Bagging ($6\%$) est bien
inférieure à celle atteinte par le meilleur des arbres précédents ($8,6\%$).

Une forêt aléatoire (construite à l'aide du package \texttt{randomForest} avec
les paramètres par défaut) atteint, quant à elle, un taux d'erreur test de $5,2\%$, 
qui a encore diminué par rapport au Bagging. Rappelons que l'idée de cette amélioration qui est quasiment
systématiquement observée, est que la collection d'arbres construite
avec les forêts aléatoires est plus diversifiée que celle construite
avec le Bagging (du fait des choix aléatoires de variables à chaque
n\oe{}ud). Cette diversité est bénéfique au moment de l'agrégation. On peut
penser aussi que l'on ``explore'' plus l'espace des prédicteurs.

\subsection{Variantes et extensions}\label{variantes-et-extensions}

\subsubsection{Variantes}

Comme mentionné ci-dessus, différentes versions des forêts 
aléatoires, de structure plus ou moins fortement dépendantes des données, ont
été proposées. On parle souvent de forêts purement aléatoires lorsque les
partitions associées aux arbres sont choisies aléatoirement indépendamment
des observations de $\mathcal{L}_n$ (\cite{breiman2000}, 
\cite{biau2008consistency}, \cite{arlot2014analysis}).
Certaines variantes ont essentiellement été développées pour l'obtention 
de résultats théoriques, mais d'autres constituent des variantes assez
compétitives sur le plan des performances avec les RF-RI (voir par exemple 
\cite{cutler_zhao_CSS2001,Geurts06,duroux2016impact}).
De surcroît, elles sont la plupart du temps
beaucoup plus simples à définir, à calculer, à mettre à jour et 
s'adaptent par exemple plus facilement au contexte des flux de données (voir la 
section \ref{forets-aleatoires-en-ligne} dédiée aux RF en ligne).

Suivant \cite{Geurts06}, introduisons la méthode Extra-Trees (pour
Extremly Randomized Trees). Le principe est, ici, de tirer aléatoirement
\(m\) variables à chaque n\oe{}ud, puis de choisir aléatoirement un point
de coupure pour chaque variable : pour une variable continue \(X^j\), on
tire le seuil \(d\) de façon uniforme dans le segment délimité par la
plus petite et la plus grande valeur de \(X^j\) des observations
présentent dans le n\oe{}ud courant. La coupure est alors
\(\{ X^j \leq d \} \cup \{ X^j > d \}\). Nous récupérons alors \(m\)
coupures, et choisissons la meilleure coupure parmi celles-ci,
c'est-à-dire celle qui minimise la fonction de coût considérée. Le choix
des \(m\) variables à chaque n\oe{}ud et le choix de la meilleure coupure
sont les mêmes que dans les RF-RI. Cependant, ici un aléa
supplémentaire est introduit au niveau du point de coupure. Là où les
RF-RI optimisent le point de coupure sur \(\mathcal{L}_n\),
les Extra-Trees tire ce point de coupure aléatoirement. \cite{Geurts06}
illustrent dans leur article les très bonnes performances des
Extra-Trees, qui apportent même parfois des améliorations par rapport à
la méthode de référence RF-RI.

Nous présentons un résumé du type de randomisation (ou aléa
supplémentaire) utilisée pour construire ces forêts aléatoires. Cette
présentation est en grande partie basée sur celle de \cite{Liu05}.

\begin{enumerate}
    \item Randomisation préalable à la construction de l'arbre :

    \begin{enumerate}
        \item Ré-échantillonnage : \textit{e.g.} Bagging, Random Forests-RI ;
        \item Tirage d'un sous-ensemble de variables (fixe pour tout l'arbre) : 
 \textit{e.g.} Random Subspace ;
        \item Perturbation des sorties :  \textit{e.g.} Randomizing Outputs.
    \end{enumerate}

    \item Randomisation au coeur de la construction de l'arbre :

    \begin{enumerate}
        \item Tirage d'un sous-ensemble de variables à chaque n\oe{}ud :  \textit{e.g.} 
Random Forets-RI, Extra-Trees ;
        \item Tirage d'une coupure parmi de bonnes coupures candidates :  \textit{e.g.} 
Randomization (\cite{dietterich2000experimental}) ;
        \item Tirage du point de coupure :  \textit{e.g.} Extra-Trees.
    \end{enumerate}

\end{enumerate}

Bien entendu, il existe pour chaque type d'aléa, plusieurs
randomisations possibles : ré-échantillonnage avec ou sans remise,
sous-échantillonné ou non, tirage uniforme ou suivant une autre loi des
sous-ensembles de variables, etc. Il est également possible de combiner
plusieurs randomisations dans une même méthode, comme c'est déjà le cas
dans Random Forets-RI ou Extra-Trees. Néanmoins, plusieurs associations
de ces randomisations ont déjà été testées par les auteurs des méthodes
citées ici. Finalement, seules les méthodes présentant les meilleurs
résultats ont été retenus. Par exemple, \cite{Geurts06} ont essayé de
coupler le Bagging et Extra-Trees, mais ont montré qu'en pratique cela
n'apporte pas d'amélioration, ceci étant sûrement dû au fait que la
randomisation des partitions est déjà assez forte dans Extra-Trees (voir
les commentaires en \ref{garantie-theorique-1}).

\subsubsection{Extensions}

Tout d'abord, on peut associer à toutes les extensions des arbres CART 
listées dans la section \ref{extensions} des extensions naturelles des forêts.

Signalons quelques-unes des extensions des forêts développées pour des objectifs variés. 
Par exemple, pour des problèmes de classement \cite{Clemencon2013},
pour l'analyse des données de survie \cite{hothorn2006survival}  et \cite{Ishwaran2008}, ou encore pour
la régression quantile \cite{Meinshausen2006} pour se limiter aux plus connues.

On trouve aussi dans \cite{yan2013cluster} une extension, plus surprenante, des RF au 
contexte de la classification non supervisée pour définir les Cluster forests. 
Un article de revue sur ce type de sujet est récemment paru visant les 
applications de type chimiométrie (voir \cite{afanador2016unsupervised}).

Suivant la remarque de Breiman, les RF conduisent à des résultats d’autant meilleurs que la diversité des arbres de la forêt est grande. Ainsi, récemment, certaines extensions ont été définies par l'amélioration d'une RF initiale.  \cite{fawagreh2015outlier}  utilisent une technique d'apprentissage non supervisé (Local Outlier Factor, LOF) pour identifier les arbres divers dans la forêt puis, ils effectuent l'élagage de l’ensemble en sélectionnant les arbres avec les scores de LOF les plus élevés pour produire une extension de RF appelé LOFB-DRF, beaucoup plus petite en taille et souvent meilleure. Ce schéma peut être étendu en utilisant d'autres mesures de diversité, voir \cite{tang2006analysis}  qui présentent une analyse théorique sur six mesures de diversité existantes. 

Une autre extension possible est de pondérer a posteriori les arbres pour améliorer la performance prédictive, voir par exemple récemment \cite{winham2013weighted}.

D'autres variantes ont été introduites pour sortir du cadre des arbres
avec des coupures parallèles aux axes, déjà par \cite{Breiman01} avec
les Random Forests-RC (RC pour "random combination"), voir aussi plus récemment
\cite{blaser2015random, menze2011oblique}.

Une très récente variante due à \cite{biau2016neural}, exploite une manière
d’associer à un arbre un réseau de neurones à deux couches cachées (la
première contient les n\oe{}uds non terminaux et la seconde les feuilles
de l’arbre, les entrées sont les variables explicatives et la sortie
la variable à expliquer) et introduit ainsi une variante neuronale des
forêts aléatoires qui hérite naturellement des techniques
d’optimisation et de réglage des paramètres des réseaux de neurones.
Dans ce contexte, des résultats de consistance sont obtenus et le gain
sur le plan des performances est substantiel tant vis-à-vis des
réseaux de neurones que des forêts aléatoires, au prix toutefois d’une
plus grande difficulté de réglage des très nombreux paramètres.

\section{Forêts aléatoires et sélection de
variables}\label{forets-aleatoires-et-selection-de-variables}

Par le passé, les problèmes considérés en statistique comportaient typiquement beaucoup
d'observations (\(n\) de l'ordre de quelques centaines ou milliers) et
peu de variables (\(p\) seulement de l'ordre de la ou de quelques dizaines). Les progrès
technologiques font que l'acquisition de données est devenue de plus
en plus facile techniquement et de nos jours des bases de données
gigantesques sont collectées ou simplement disponibles, quasi-quotidiennement. Les techniques
statistiques classiques ne suffisent plus pour traiter ces nouvelles
données. Le nombre de variables \(p\), peut désormais atteindre des
dizaines voire des centaines de milliers. Dans le même temps, pour
beaucoup d'applications, le nombre d'observations \(n\), se trouve
réduit à quelques dizaines. Le domaine typique de telles situations est
le domaine biomédical où l'on peut maintenant faire énormément de
mesures sur un individu donné (mesures d'expression de gènes par
exemple), mais le nombre d'individus sur lequel l'expérience est menée est
réduit (ainsi dans le cas d'étude d'une maladie, le nombre de porteurs de la
maladie qui participent à une étude est souvent limité).

Dans la suite, nous dirons que les données considérées sont de grande
dimension lorsque le nombre de variables est
très grand devant le nombre d'observations \(n<<p\). Nous avons en tête des
problèmes où \(n\) est de l'ordre de \(100\) et \(p\) de l'ordre de
plusieurs milliers. Un des avantages des forêts aléatoires est qu'elles sont 
très performantes aussi bien pour des problèmes classiques (où $p \leq n$) que pour des
problèmes de grande dimension.

Dans de nombreux problèmes, et en particulier dans le cas de données de
grande dimension, en plus d'un bon prédicteur, les
praticiens souhaitent également avoir des informations supplémentaires
sur les variables du problème et connaître les variables effectivement
utiles pour expliquer le lien entrée-sortie. Ils désirent donc que le
statisticien leur propose une sélection des variables. De plus, dans ce cadre, 
il est naturel de penser que
relativement peu de variables (disons au maximum de l'ordre de \(n\))
agissent réellement sur la sortie et il convient de faire des hypothèses supplémentaires (de sparsité)
pour lui redonner un sens. On trouve dans \cite{gira_2014}
une très complète présentation des problèmes et techniques mathématiques 
pour aborder ce type de questions.

On peut en citer quelques méthodes de sélection de variables pour des données
de grande dimension.

Commençons par une étude empirique dans \cite{Poggi06} où l'on introduit une
méthode basée sur le score d'importance des variables fourni par
l'algorithme CART. Dans un état d'esprit proche, citons aussi \cite{questier2005use}. 

En considérant le problème plus généralement, 
\cite{Guyon02,Rakotomamonjy03,BenIshak08} utilisent le score calculé par
les Support Vector Machines (SVM : \cite{Vapnik00}). \cite{Diaz06}
proposent une procédure de sélection de variables basée sur l'importance
des variables des forêts aléatoires. Toutes ces méthodes
calculent tout d'abord un score pour chacune des variables, puis
procèdent à une introduction séquentielle de variables (méthodes
``forward'), ou une élimination séquentielle de variables
(méthodes ''backward`` ou RFE pour Recursive Feature Elimination), ou exécutent des méthodes pas-à-pas
(méthodes ''stepwise``) mêlant introduction et élimination de
variables.

On trouve dans \cite{Fan08} une méthode en deux
temps : une première étape d'élimination de variables pour atteindre une
situation raisonnable où \(p\) est de l'ordre de \(n\), puis une
deuxième étape de type forward basée par exemple sur le Least Absolute
Shrinkage and Selection Operator (Lasso, \cite{tibshirani_JRSSB1996}).

Dans cet esprit, un schéma général pour calculer un score d'importance pour les
variables  est proposé dans \cite{LeCao07}, puis les auteurs utilisent ce schéma 
avec comme méthode de base CART et
SVM. Leur idée est d'apprendre un vecteur de poids sur toutes les
variables (leur meta-algorithme se nomme Optimal Feature Weighting
(OFW)) : une variable avec un poids fort est importante, tandis qu'une
variable avec un poids faible est inutile.

Enfin, plus récemment, des méthodes d'amélioration du Lasso, pour la sélection de
variables, ont été mises au point. Ces dernières ont des points communs
avec les méthodes d'ensemble. En effet, au lieu de chercher à
faire de la sélection  ``en un coup''  avec un Lasso classique, elles
cherchent à construire plusieurs sous-ensembles de variables et à les
mettre ensuite en commun. Dans Bolasso (pour Bootstrap-enhanced Lasso),
introduit par \cite{Bach08}, on génère plusieurs échantillons bootstrap
puis on lance sur chacun d'eux la méthode du Lasso. Bolasso est donc à
mettre en parallèle avec la méthode du Bagging de \cite{Breiman96}. Dans
Randomized Lasso, \cite{Meinshausen10} choisissent de générer plusieurs
échantillons par sous-échantillonnage et rajoutent un aléa
supplémentaire dans la construction même du Lasso. Randomized Lasso est
donc lui à rapprocher des forêts aléatoires Random Forests-RI. Dans le même esprit, on 
peut mentionner aussi \cite{fellinghauer2013stable} qui utilisent les RF 
pour l'estimation robuste des modèles graphiques.

L'intérêt pour le sujet reste vif, comme en témoigne la référence récente 
\cite{Hapfelmeier2013} qui
propose une nouvelle approche de sélection utilisant les forêts aléatoires ou
encore
\cite{Cadenas13} qui décrit et compare ces différentes approches, dans un
papier de synthèse.

Dans la suite de cette section, on définit tout d'abord l'importance des variables 
par permutation avant de présenter une procédure de sélection des variables
de combinant étroitement scores d'importances, forêts aléatoires et des 
stratégies usuelles de sélection. Ensuite on donne quelques liens vers des 
résultats théoriques partiels avant d'appliquer la procédure au données \ctt{spam}
et de mentionner variantes et extensions.

\subsection{Importance des variables}\label{importance-des-variables}

Lorsqu'il ne s'agit pas seulement d'éliminer des variables du modèle ou de 
retenir seulement des variables explicatives suffisantes pour obtenir une 
bonne prédiction, on peut être intéressé par construire une hiérarchie de 
toutes les variables fondée sur l'importance vis-à-vis de la variable réponse.
Un tel indice d'importance  fournit donc un classement des variables, 
de la plus importante à la moins importante. 

On trouvera dans \cite{Aze03} une discussion générale sur l'importance des variables.
Il est clair que cette notion a été relativement peu examinée par les statisticiens et principalement 
dans le cadre des modèles linéaires, comme en atteste le papier de synthèse de
\cite{gromping2015variable} ou la récente thèse de \cite{wallard2015analyse}. 
Il se trouve que les forêts aléatoires offrent un cadre idéal, alliant
une méthode non-paramétrique,  ne prescrivant pas de forme particulière à la relation entre $Y$ et les
composantes de $X$, au ré-échantillonnage, pour disposer d'une définition à la fois efficace et 
commode de tels indices.

Dans ce cadre, l'une des mesures largement utilisées de
l'importance d'une variable donnée est l'accroissement moyen de l'erreur
d'un arbre dans la forêt lorsque 
les valeurs observées de cette variable sont permutées au hasard dans les
échantillons OOB. 

\noindent
Fixons \(j \in \{1,\ldots,p\}\) et détaillons le calcul de l'indice
d'importance de la variable \(X^j\). Considérons un échantillon
bootstrap \(\mathcal{L}_n^{\Theta_l}\) et l'échantillon
\(\mathrm{OOB}_l\) associé, c'est-à-dire l'ensemble des observations qui
n'apparaissent pas dans \(\mathcal{L}_n^{\Theta_l}\). Calculons
\(\mathrm{errOOB}_l\), l'erreur commise sur \(\mathrm{OOB}_l\) par
l'arbre construit sur \(\mathcal{L}_n^{\Theta_l}\) (erreur quadratique
moyenne en régression, proportion de mal classés en classification).
Permutons alors aléatoirement les valeurs de la \(j\)-ième variable dans
l'échantillon \(\mathrm{OOB}_l\). Ceci donne un échantillon perturbé,
noté \(\widetilde{\mathrm{OOB}_l}^j\). Calculons enfin
\(\mathrm{err}\widetilde{\mathrm{OOB}_l}^j\), l'erreur sur l'échantillon
\(\widetilde{\mathrm{OOB}_l}^j\). Nous effectuons ces opérations pour
tous les échantillons bootstrap. L'importance de la variable \(X^j\),
\(\mathrm{VI}(X^j)\), est définie par la différence entre l'erreur
moyenne d'un arbre sur l'échantillon OOB perturbé et celle sur
l'échantillon OOB :

\[\mathrm{VI}(X^j) = \frac{1}{q} \sum_{l=1}^q \left( 
\mathrm{err}\widetilde{\mathrm{OOB}_l}^j - \mathrm{errOOB}_l \right)\;.\]

Ainsi, plus les permutations aléatoires de la \(j\)-ième variable
engendrent une forte augmentation de l'erreur, plus la variable est
importante. A l'inverse, si les permutations n'ont quasiment aucun effet
sur l'erreur (voire même la diminuent ce qui fait que VI peut être légèrement 
négative), la variable est considérée comme très peu
importante. 

Précisons que ce calcul de l'importance est exactement celui qui est
exécuté dans le paquet \texttt{randomForest}. La définition qui est
donné de l'importance dans \cite{Breiman01} est légèrement différente.
Les échantillons OOB perturbés sont obtenus de la même manière. Par
contre, l'importance d'une variable est alors définie par la différence
entre l'erreur OOB sur les échantillons OOB perturbés, et l'erreur OOB
initiale. Nous pensons que la raison de ce changement est que dans le
calcul de l'erreur OOB, il y a une étape d'agrégation. L'agrégation a
tendance à amoindrir les erreurs des arbres individuels, donc
l'utilisation de l'erreur OOB dans le calcul tend à atténuer les effets
des permutations aléatoires des variables. Or, dans le calcul de
l'importance, nous voulons au contraire, que les effets de ces
permutations soient les plus visibles possible. C'est pourquoi l'erreur
OOB est remplacée par l'erreur moyenne sur tous les arbres, dans
laquelle aucune étape d'agrégation n'est effectuée.

Des études, à l'aide de simulations pour l'essentiel, ont été menées pour
illustrer le comportement de l'importance des variables des forêts
aléatoires, en particulier dans les travaux de
\cite{Str07, Strobl08, Archer08, Gen08, Gen10, Gre13}. 
Dans ce type d'études empiriques, il est souvent  difficile de tirer des
conclusions générales très assurées, on peut néanmoins noter quelques 
comportements intéressants de ce score de VI pour la sélection :

\begin{itemize}
\item Les variables hors modèle, ou non informatives, ont une VI très faible.
\item La variabilité au travers des répétitions des VI des variables hors modèle est moins importante
que celle des variables informatives.
\item En présence d'un groupe de variables très corrélées des effets de masquage 
apparaissent mais demeurent limités sauf lorsque le nombre de variables du groupe devient 
dominant par rapport aux autres variables informatives.
\item Une évaluation stable de ces VI requiert des forêts touffues ($ntree$ grand)
et quelques répétitions de telles forêts sont utiles pour en quantifier la variabilité.
\end{itemize}

\bigskip

Nous allons donc nous concentrer sur la performance de prédiction des RF en
se restreignant à l'estimation par l'erreur out-of-bag (OOB)  et 
sur la quantification de l'importance des variables par permutation 
qui sont  les ingrédients clés pour notre stratégie de sélection des variables. 

\subsection{Sélection de variables}\label{selection-de-variables}

Nous proposons dans \cite{Gen10} (voir aussi \cite{genuer2015vsurf} pour 
le package correspondant) une méthode de sélection de variables qui est 
une procédure ``automatique'' au sens où il
n'y a aucun \textit{a priori} à apporter pour faire la sélection. Par exemple, il
n'est pas nécessaire de préciser le nombre de variables que l'on
souhaite obtenir : la procédure s'adapte aux données pour fournir le
sous-ensemble de variables final. De plus, notre méthode procède en deux
étapes : la première (assez grossière et descendante) consiste à seuiller sur
l'importance des variables dans le but d'éliminer un grand nombre de
variables inutiles, tandis que la seconde (plus fine et ascendante) consiste en une
introduction de variables dans des modèles de forêts aléatoires.

Nous  distinguons en outre deux objectifs  de sélection de variables, que nous appellerons
(bien que cette terminologie puisse prêter à confusion) interprétation et prédiction :

\begin{enumerate}
    \item Pour un but d'interprétation, nous cherchons à sélectionner toutes les 
variables $X^j$ fortement reliées à la variable réponse $Y$ (même si les 
variables $X^j$ sont corrélées entre elles).
    \item Alors que pour un but de prédiction, nous cherchons à sélectionner un 
petit sous-ensemble de variables suffisant pour bien prédire la variable 
réponse.
\end{enumerate}

Typiquement, un sous-ensemble construit pour satisfaire (1) pourra
contenir beaucoup de variables, qui seront potentiellement très
corrélées entre elles. Au contraire, un sous-ensemble de variables
satisfaisant (2) contiendra peu de variables, avec très peu de
corrélation entre elles.

Une situation typique illustre la distinction entre les deux types
de sélection de variables. Considérons un problème de classification en grande 
dimension ($n<<p$)
pour lequel chaque variable explicative est associée à un pixel dans une image 
ou un voxel dans une 
image 3D comme dans les problèmes de classification en activité cérébrale 
(IRMf), voir par exemple
\cite{Gen10b}. Dans
de telles situations, il est supposé que beaucoup de variables sont inutiles ou 
non informatives
et qu'il existe des groupes inconnus de prédicteurs fortement corrélés
correspondant à des régions du cerveau impliquées dans la réponse à une 
stimulation donnée. Même si les deux objectifs de  sélection des variables peuvent
être d'intérêt, il est manifeste que trouver toutes les variables importantes hautement
reliée à la variable réponse est utile pour l'interprétation, puisqu'il
correspond à la détermination des régions entières du cerveau ou
d'une image. Bien sûr, la recherche d'un petit nombre de variables
suffisantes pour une bonne prédiction, permet d'obtenir les variables les plus 
discriminantes dans les régions précédemment mises en évidence mais est moins 
prioritaire dans ce contexte. Pour une approche plus formelle
de cette distinction, on peut se reporter à l'intéressant papier de 
\cite{Nil07}.

\bigskip

Notre méthode de sélection de variables tente donc de satisfaire les deux
objectifs précédents. Décrivons un peu plus précisément notre 
procédure en deux étapes qui est basée sur des forêts aléatoires
comportant un très grand nombre d'arbres (typiquement \(ntree=2000\)).
\medskip

\begin{itemize}
\item Etape 1. Elimination préliminaire et classement :
\medskip
  \begin{itemize}
  \item Classer les variables par importance décroissante (en fait par VI 
moyenne sur
   $50$ forêts typiquement). 
\medskip
  \item Eliminer les variables de faible importance (soit $m$ le nombre de 
  variables conservées).
  \medskip

  {\footnotesize Plus précisément, en partant de cet ordre, on considère la suite ordonnée des 
  écart-types des VI correspondants (sd) que l'on utilise pour estimer la valeur 
d'un seuil
  sur les VI. Comme la variabilité des VI plus grande pour les vraies variables 
du 
  modèle que pour les variables non informatives, la valeur du seuil est donnée 
par l'estimation 
  de l'écart-type de VI pour ces dernières variables. Ce seuil est fixé au 
minimum
  prédit par le modèle CART ajusté aux données $(X,Y)$ où les $Y$ sont les sd 
des VI 
  et les $X$ sont les rangs.

Ensuite seules les  variables dont l'importance
 VI moyenne est plus grande que ce seuil sont gardées.}
    
  \end{itemize}
\medskip
  
\item Etape 2. Sélection de variables :
\medskip

  \begin{itemize}
  
  \item Pour l'\emph{interprétation} : on construit la collection de modèles 
emboîtés 
  constituées par les forêts construites sur les  $k$ premières variables, pour 
$k=1$ à $m$ et
    on sélectionne les variables du modèle conduisant à la plus faible erreur 
    OOB. Ceci conduit à considérer $m^{\prime}$ variables.
 \medskip
   
  {\footnotesize Plus précisément, on calcule les moyennes (typiquement sur 25 forêts) des erreurs OOB des modèles emboîtés en commençant par celui ne comportant que la variable la plus importante et se terminant par celle impliquant toutes les variables importantes retenues précédemment. Idéalement, les variables du modèle conduisant à l'erreur  OOB la plus faible sont choisies. En fait, pour faire face à l'instabilité, nous utilisons une astuce classique : nous sélectionnons  le plus petit modèle avec une erreur inférieure à la plus faible erreur OOB augmentée de l'estimation de l'écart-type de cette erreur (basée sur les 25 mêmes répétitions de forêts).
  
  }
\medskip
    
  \item Pour la \emph{prédiction} : à partir des variables retenues pour 
l'interprétation, on cons- truit une suite de modèles en introduisant 
séquentiellement, dans l'ordre d'importance croissant, et en testant  
itérativement les variables. Les variables du dernier modèle sont sélectionnées.
\medskip
    
    {\footnotesize Plus précisément, l'introduction séquentielle des variables est basée sur le 
test suivant :
    une variable est ajoutée seulement si l'erreur OOB décroît plus qu'un seuil. 
L'idée est que
  l'erreur OOB doit décroître plus que la variation moyenne engendrée par 
l'ajout de variables non informatives.
    Le seuil est fixé à la moyenne des valeurs absolues des différences 
premières des erreurs OOB entre le 
    le modèle à $m^{\prime}$ variables et celui à $m$ variables :

\begin{equation}\label{mean_jump}
 \frac{1}{m - m^{\prime}} \sum_{j=m^{\prime}}^{m-1} \left| \, errOOB(j+1) - 
errOOB(j) \, \right|
\end{equation}

où $errOOB(j)$ est l'erreur OOB de la  forêt construite grâce aux $j$ variables 
les
plus importantes.}

  \end{itemize}
\end{itemize}

Il faut noter que l'importance utilisée ici n'est pas normalisée, comme cela est
souvent le cas. En effet, au lieu de considérer comme mentionné 
dans \cite{BreiCutl04}, que les importances  brutes sont indépendantes et de même loi, de les
normaliser et supposer leur normalité asymptotique, on préfère ici une solution 
entièrement 
pilotée par les données. C'est une des clés de notre stratégie qui préfère estimer 
directement la variabilité
au travers des répétitions de forêts plutôt que d'utiliser la normalité quand 
 \(ntree\) est suffisamment grand, qui est valide dans des conditions 
spécifiques,
en fait très difficiles à vérifier puisque dépendant lourdement  des paramètres 
de 
réglage de la méthode et des spécificités du problème.  Une normalisation dirigée
par les données nous prévient contre un comportement asymptotique mal spécifié.

Naturellement, un tel schéma peut être modifié et adapté dans diverses directions, par exemple
à la marge, en réévaluant l'ordre des variables à chacune des étapes ou encore plus profondément, 
en classant les variables d'origine et en enlevant  la moins importante puis en réitérant pour procéder 
à la phase d'élimination ou encore, en explorant plus de configurations suivant la valeur de $m$.

\subsection{Garantie théorique}\label{garantie-theorique-2}

Il est très difficile d'obtenir des résultats de consistance de la sélection pour 
ce type de méthode dans un contexte de grande dimension. 
En général, on  dispose de résultats sur des méthodes 
de même inspiration, essentiellement sur le modèle linéaire (voir \cite{chen2014variable},
\cite{hesterberg2008least}, \cite{paul2008preconditioning} ou le livre de \cite{gira_2014}) et l'on
utilise des heuristiques variées pour fouiller très partiellement la famille 
sous-jacente de modèles.

Un élément utile, même s'il est modeste, pour asseoir la validité théorique de ce type de 
procédure de sélection de variables basée sur une méthode progressive classique 
combinée avec l'utilisation de la mesure d'importance VI, est qu'une variable non informative, 
c'est-à-dire non incluse dans le modèle sous-jacent, a une importance théorique nulle.

\cite{Ishwaran07} présente une première approche pour tenter d'expliquer
de façon théorique le comportement de l'importance des variables, en
étudiant une version simplifiée cette importance.

Récemment, \cite{Gre13} établit théoriquement que, dans le cas des modèles 
additifs,
les variables non pertinentes ont une importance VI théorique nulle. Plus
précisément, on dispose dans ce cas particulier d'une expression explicite de l'importance d’une variable 
comme la variance de la composante associée,  à un facteur 2 près. Une conséquence immédiate, 
à l’évidente portée pratique, est qu'une variable non informative ou hors modèle a une importance nulle. 
Toujours dans ce cadre, une étude théorique de l’impact de la corrélation entre prédicteurs sur l’importance 
est menée. Fort de ces deux aspects, une variante de l’algorithme RFE (Recursive Feature Elimination) 
reposant sur des classements des variables par importance, est proposée, puis étudiée par simulation 
et enfin, appliquée sur des données réelles d’aviation.

Notons enfin qu'un résultat similaire pour l'indice d'importance basé sur la diminution moyenne de l'impureté  
(voir \ref{variantes-des-mesures-dimportance}) a été prouvée dans \cite{louppe2013understanding} et utilisée pour définir une stratégie de
sélection de variables dans une situation de variables dont les interactions dépendent du contexte 
(voir \cite{sutera2016context}).

\subsection{Exemple}\label{exemple-2}

\begin{figure}[!ht]
\begin{center}
\includegraphics[width=0.9\textwidth]{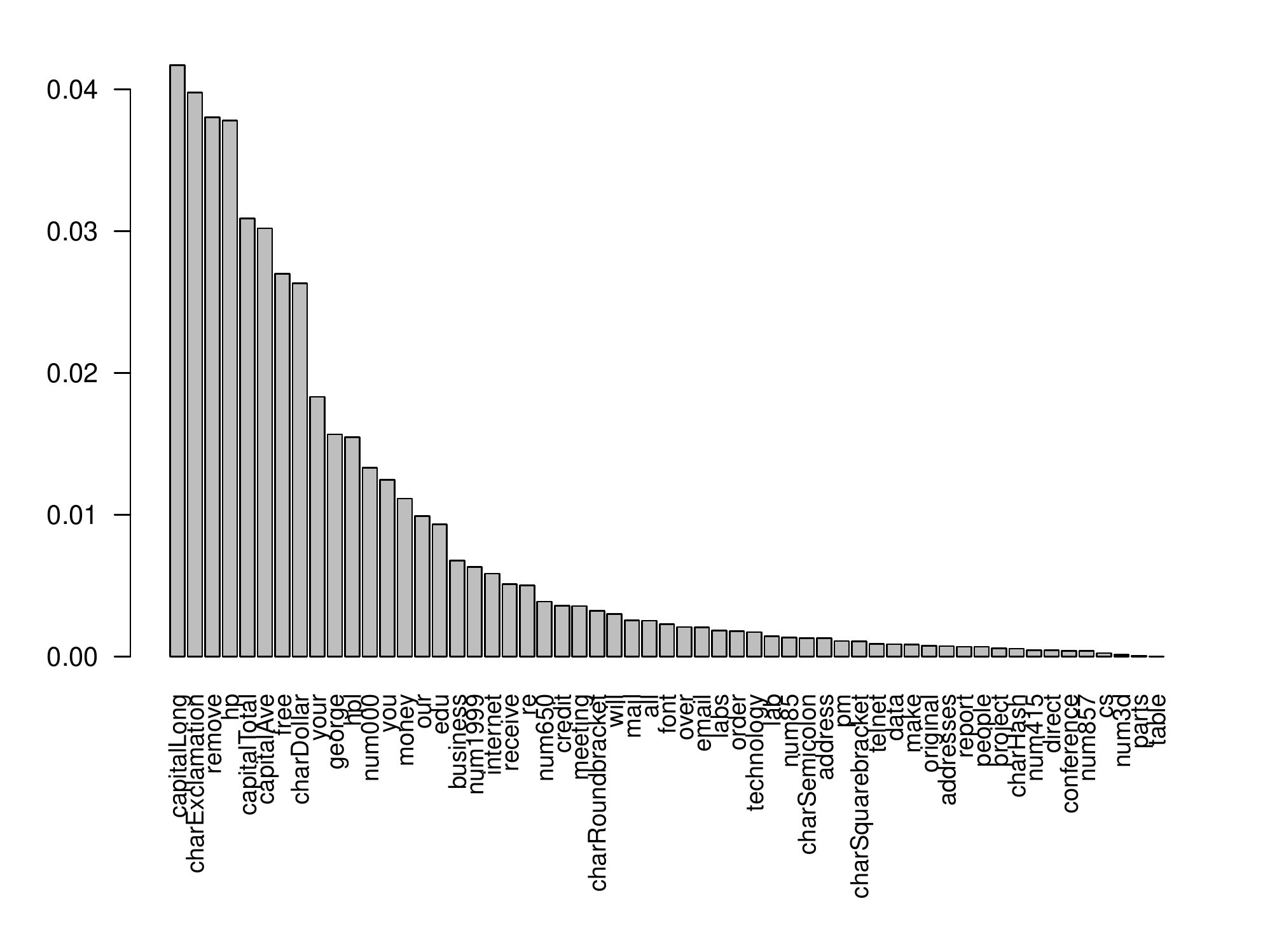}
  \caption{Importance des variables en ordre décroissant}
  \label{VI_decroissantes}
\end{center}
\end{figure}

Sur l'exemple des données \ctt{spam}, le classement des variables par ordre 
décroissant d'importance conduit à la 
Figure \ref{VI_decroissantes}.

En se restreignant aux huit plus importantes qui semblent se dégager, on trouve les
proportions d'occurrences des mots ou caractères \ctt{remove}, \ctt{hp}, \ctt{\$}, \ctt{!}, \ctt{free} 
ainsi que 
les 3 variables liées aux longueurs des suites de lettres majuscules. Ce qui est 
assez conforme à l'intuition.

Remarquons ici que les variables impliquées dans les premières découpes de 
l'arbre optimal (voir la Figure \ref{ArbreCARTelague}) ne sont pas en tête du 
classement par importance des variables,
même si elles apparaissent bien classées. Au contraire, certaines variables
peuvent être importantes et ne pas apparaître dans l'arbre en question, tel est 
le cas ici de la variable la plus importante : \ctt{capitalLong}. 

\begin{figure}[!ht]
\begin{center}
\includegraphics[width=0.9\textwidth]{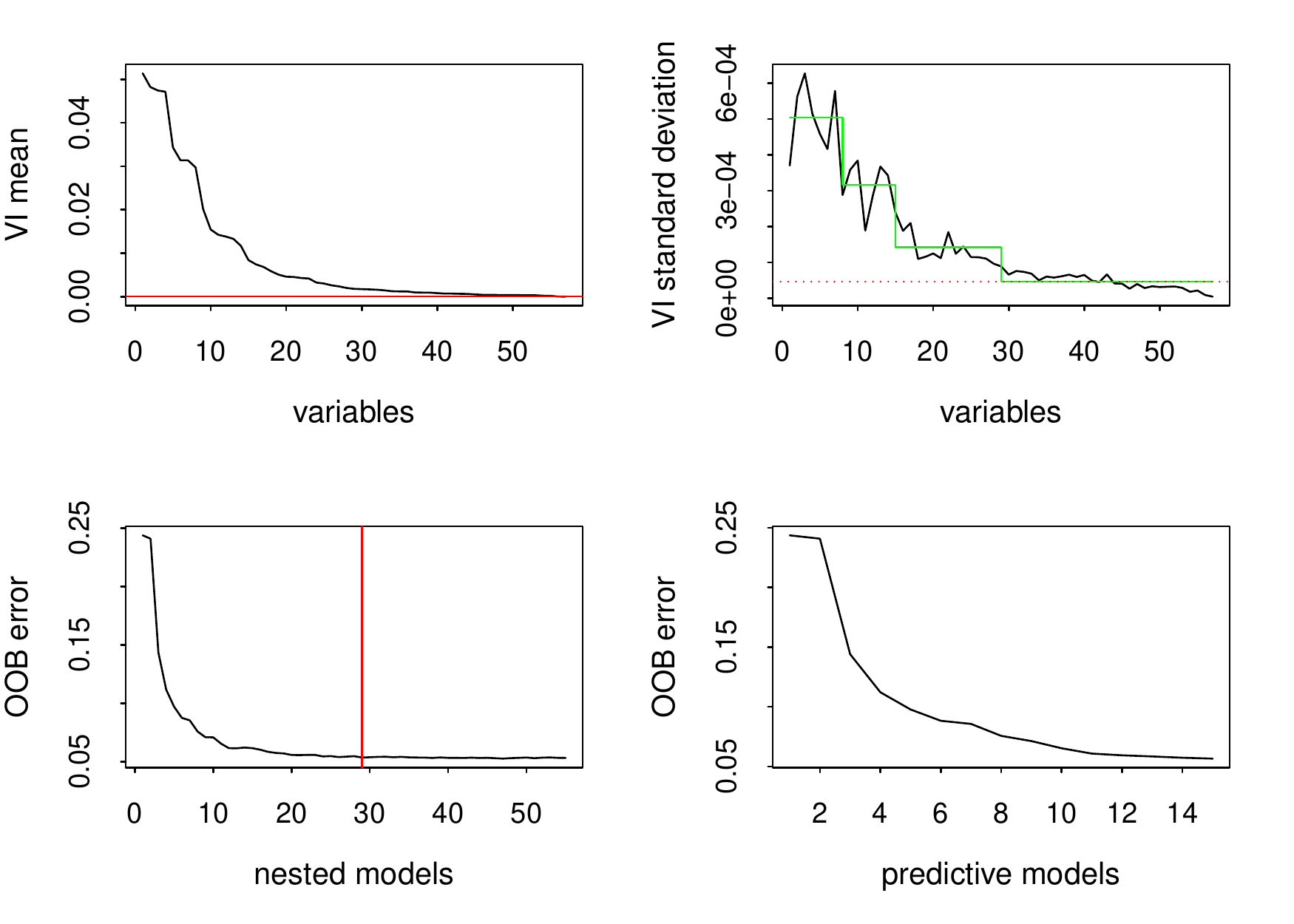}
  \caption{La procédure \ctt{VSURF} appliquée aux données \ctt{spam}. Les graphiques du haut résument l'étape de seuillage, celui d’en bas à gauche illustre l’obtention de l’ensemble d’interprétation et en bas à droite, de prédiction.}
    \label{VSURF_4plots}
\end{center}
\end{figure}

On peut ensuite s'intéresser au problème de la sélection des deux ensembles 
de variables pour l'interprétation et la prédiction dont l'objectif est de sélectionner
deux sous-ensembles beaucoup plus petits que les 57 variables initiales sans trop dégrader
les performances de la forêt initiale.
Les résultats de l'application du package \ctt{VSURF}
(voir \cite{genuer2015vsurf}) sur les données \ctt{spam} sont rassemblés
dans la Figure \ref{VSURF_4plots}.

Partant de $57$ variables, la première étape d'élimination (voir les deux graphiques du haut de la Figure) les conserve quasiment toutes, 
l'étape d'interprétation (voir le graphique en bas à gauche) sélectionne environ la moitié ($29$) des variables et pour la prédiction (voir le graphique en bas à droite)
une quinzaine. 

\begin{table}[!ht]
  \centering
  \begin{tabular}{c|ccc}
    Forêt & initiale & interprétation & prédiction \\
    \hline
    Erreur test & 0.052 & 0.056 & 0.060
  \end{tabular}
  \caption{Erreurs test des RF utilisant respectivement les variables sélectionnées après l'étape d'interprétation et celle de prédiction, comparées à l'erreur d'une RF dite initiale, construite avec toutes les variables}
  \label{perfos_vsurf}
\end{table}

Les performances sont rassemblées dans la Table \ref{perfos_vsurf}. Il faut noter que, dans cet exemple, il y a une dégradation de l'erreur test vis-à-vis de la forêt initiale : de 
l'ordre de 8\% pour l'ensemble d'interprétation et à nouveau 8\% pour l'ensemble de prédiction. Mais elle est limitée
si l'on remarque que l'erreur test finale est obtenue avec moins de 15 variables au lieu des 57 initialement considérées,
et qu'elle est de l'ordre de grandeur de celle obtenue par le bagging.

\subsection{Variantes et extensions}\label{variantes-et-extensions-1}

\subsubsection{Variantes des mesures
d'importance}\label{variantes-des-mesures-dimportance}

Notons que \cite{Str07} constatant que les VI sont biaisées en faveur 
des variables corrélées, propose dans   \cite{Strobl08} une variante du schéma 
de permutation
appelé importance conditionnelle ne présentant pas ce défaut mais qui en 
augmente très 
notablement le coût de calcul. Il faut noter que ceci 
semble être particulièrement critique pour les problèmes de grande dimension 
avec
des prédicteurs fortement corrélés. Néanmoins, nos expériences précédentes
\cite{Gen10} sur la sélection de variables et, plus récemment,
l'étude théorique de \cite{Gre13} montrent que, dans certaines situations, au 
contraire
les scores de VI sont biaisés en faveur des variables non corrélées.

Signalons à nouveau qu'un indice basé sur la diminution moyenne de l'impureté 
(Mean Decrease Impurity importance (MDI)) et dont l'inspiration est dans le droit fil de l'importance des 
variables définie dans la méthode CART existe aussi. Il est défini, suivant \cite{Breiman01} ainsi : pour une variable donnée, son MDI
est la somme des diminutions de l'impureté pour tous les n\oe{}uds où la variable est utilisée dans la découpe,
pondérée par la proportion d'individus dans le n\oe{}ud, en moyenne sur tous les arbres de la forêt.
Même si son utilisation semble moins fréquente, elle a donné lieu à des 
résultats théoriques dans \cite{louppe2013understanding} et trouve des applications pratiques 
 intéressantes malgré des limitations réelles comme la forte dépendance aux 
paramètres de la forêt et le fait qu'elle soit biaisée pour les arbres non élagués.

\subsubsection{Extension à la sélection de variables
fonctionnelles}\label{extension-a-la-selection-de-variables-fonctionnelles}

Dans \cite{gregorutti2015grouped}, on trouve une extension directe de l’importance des variables par permutation à un 
groupe de variables, il suffit pour cela de permuter, dans les échantillons OOB intervenant dans la définition de VI (voir 
\ref{importance-des-variables}), non pas seulement une variable
mais de permuter solidairement toutes les variables d'un même groupe. Pour des modèles additifs fonctionnels, les résultats
théoriques de \cite{Gre13} sont étendus.
Ensuite une stratégie de sélection de variables dans un cadre de données fonctionnelles multivariées s'en déduit
par le biais de la sélection de groupes de variables constitués des ensembles de coefficients d'ondelettes représentant les 
différentes variables fonctionnelles et l'utilisation d'une stratégie de type RFE.

\section{Forêts aléatoires et Big
Data}\label{forets-aleatoires-et-big-data}

On ne saurait imaginer clore cet article sans évoquer le Big Data qui constitue l'un des prochains défis majeurs de la statistique. En effet, 
de récentes réflexions explorent déjà les nombreuses conséquences de
ce nouveau contexte tant du point de vue algorithmique que sur le plan
des implications théoriques (voir \cite{fan_etal_NSR2014,hoerl_etal_WIRCS,jordan_B2013}). 
 
On se contente ici, en suivant \cite{genu-2015}), de pointer quelques 
aspects spécifiques aux forêts aléatoires dans ce contexte, puisqu'elles 
y sont en principe fort bien adaptées.  

On insiste toujours lorsque l'on parle du Big Data sur l'aspect données massives, 
mais il faut rappeler qu'il est souvent associé à deux autres traits importants : les 
flux de données (\cite{aggarwal2007data}) et l'hétérogénéité
des données (voir \cite{besse_etal_p2014} pour une  introduction générale).
Le problème posé par cette grande quantité de données est double : d'abord, la 
mise en oeuvre de nombreuses procédures statistiques peut prendre trop de temps pour 
fournir des résultats acceptables et d'autre part, elle requiert de distribuer les tâches entre 
les coeurs d'un même ordinateur ou bien encore de les distribuer au travers de plusieurs
ordinateurs.

\subsection{Passage à l'échelle}\label{passage-a-lechelle}

\begin{figure}[!ht]
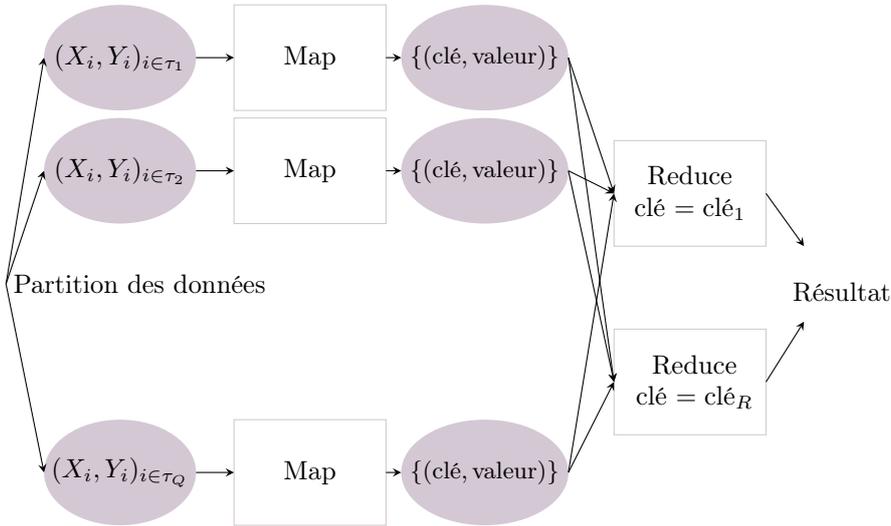

  \centering
    \begin{pgfpicture}{2 cm}{0.5 cm}{14 cm}{8 cm}
  	\color{nathgrey}
		\pgfellipse[fill]{\pgfxy(3.5,7)}{\pgfxy(1,0)}{\pgfxy(0,0.7)}
		\pgfellipse[fill]{\pgfxy(3.5,5.5)}{\pgfxy(1,0)}{\pgfxy(0,0.7)}
		\pgfellipse[fill]{\pgfxy(3.5,1.5)}{\pgfxy(1,0)}{\pgfxy(0,0.7)}
		\pgfellipse[fill]{\pgfxy(8.3,7)}{\pgfxy(1.1,0)}{\pgfxy(0,0.7)}
		\pgfellipse[fill]{\pgfxy(8.3,5.5)}{\pgfxy(1.1,0)}{\pgfxy(0,0.7)}
		\pgfellipse[fill]{\pgfxy(8.3,1.5)}{\pgfxy(1.1,0)}{\pgfxy(0,0.7)}
		
		\pgfrect[stroke]{\pgfxy(5,6.3)}{\pgfxy(2,1.4)}
		\pgfrect[stroke]{\pgfxy(5,4.8)}{\pgfxy(2,1.4)}
		\pgfrect[stroke]{\pgfxy(5,0.8)}{\pgfxy(2,1.4)}
		\pgfrect[stroke]{\pgfxy(10,2)}{\pgfxy(2,1.4)}
		\pgfrect[stroke]{\pgfxy(10,4.5)}{\pgfxy(2,1.4)}

		\color{black}
		\pgfputat{\pgfxy(2.1,4)}{\pgfbox[left,center]{Partition des données}}
		\pgfputat{\pgfxy(3.5,7)}{\pgfbox[center,center]{$(X_i,Y_i)_{i \in \tau_1}$}}
		\pgfputat{\pgfxy(3.5,5.5)}{\pgfbox[center,center]{$(X_i,Y_i)_{i \in 
\tau_2}$}}
		\pgfputat{\pgfxy(3.5,1.5)}{\pgfbox[center,center]{$(X_i,Y_i)_{i \in 
\tau_Q}$}}
		\pgfputat{\pgfxy(6,7)}{\pgfbox[center,center]{Map}}
		\pgfputat{\pgfxy(6,5.5)}{\pgfbox[center,center]{Map}}
		\pgfputat{\pgfxy(6,1.5)}{\pgfbox[center,center]{Map}}
		\pgfputat{\pgfxy(8.3,7)}{\pgfbox[center,center]{\small 
$\{(\mbox{clé},\mbox{valeur})\}$}}
		\pgfputat{\pgfxy(8.3,5.5)}{\pgfbox[center,center]{\small 
$\{(\mbox{clé},\mbox{valeur})\}$}}
		\pgfputat{\pgfxy(8.3,1.5)}{\pgfbox[center,center]{\small 
$\{(\mbox{clé},\mbox{valeur})\}$}}
		\pgfputat{\pgfxy(11,2.7)}{\pgfbox[center,center]{
		\begin{minipage}{2 cm}
			\centering Reduce\\ clé $=\mbox{clé}_R$
		\end{minipage}}}
		\pgfputat{\pgfxy(11,5.2)}{\pgfbox[center,center]{\begin{minipage}{2 cm}
			\centering Reduce\\ clé $=\mbox{clé}_1$
		\end{minipage}}}
		\pgfputat{\pgfxy(13,3.9)}{\pgfbox[center,center]{Résultat}}
		
		\pgfsetarrowsend{stealth}
		\pgfxyline(2,4)(2.5,7)
		\pgfxyline(2,4)(2.5,5.5)
		\pgfxyline(2,4)(2.5,1.5)
		\pgfxyline(4.5,7)(5,7)
		\pgfxyline(4.5,5.5)(5,5.5)
		\pgfxyline(4.5,1.5)(5,1.5)
		\pgfxyline(7,7)(7.2,7)
		\pgfxyline(7,5.5)(7.2,5.5)
		\pgfxyline(7,1.5)(7.2,1.5)
		\pgfxyline(9.4,7)(10,5.2)
		\pgfxyline(9.4,5.5)(10,5.2)
		\pgfxyline(9.4,1.5)(10,5.2)
		\pgfxyline(9.4,7)(10,2.7)
		\pgfxyline(9.4,5.5)(10,2.7)
		\pgfxyline(9.4,1.5)(10,2.7)
		\pgfxyline(12,5.2)(12.5,4.5)
		\pgfxyline(12,2.7)(12.5,3.5)
	\end{pgfpicture}
    \caption{Le paradigme de programmation MapReduce (MR)}
    \label{fig::mapreduce}
\end{figure}

Une approche pour faire face à de telles données de grande taille est de 
 ``diviser pour régner''.
Le problème d'intérêt est divisé en sous-problèmes plus simples et les solutions 
sont ensuite combinées pour résoudre le problème d'origine.
En particulier, le paradigme de programmation MapReduce (MR) (voir
Figure \ref{fig::mapreduce}) divise les données en petits
sous-échantillons, tous traités indépendamment en parallèle. Plus précisément, 
MR
se déroule en deux étapes : dans une première étape, appelée l'étape Map, 
l'ensemble de données
 est divisé en plusieurs petits blocs de données,
\((x_i,y_i)_{i\in \tau_k}\), avec \(\cup_k \tau_k = \{1,\ldots,n\}\) et
\(\tau_k\cap \tau_{k'} = \emptyset\), 
chacun d'eux étant traité séparément par exemple par un des coeurs de l'une 
des machines du cluster le cas échéant. Ces différentes tâches Map 
sont indépendantes et produisent une liste de couples de la forme 
\((\mbox{clé}, \mbox{valeur})\), où
``clé'' est une clé indexant les données qui ont été utilisées pour construite ``valeur''.
Puis, dans une seconde étape, appelée Reduce, chaque tâche Reduce traite
toutes les sorties des tâches Map correspondant à une valeur de clé donnée.

On montre (voir \cite{chu_etal_NIPS2010}) que nombre de 
méthodes statistiques
peuvent facilement s'adapter à ce paradigme : tant que la méthode est
basée sur le calcul d'une somme, des sommes partielles peuvent être calculées 
par les 
différentes tâches Map et l'étape Reduce consiste simplement dans le calcul de 
la somme finale
comme somme des sommes partielles. Cette approche peut être appliquée 
directement à la régression 
linéaire, l'analyse discriminante, la classification naïve bayésienne  et 
l'estimateur obtenu
est exactement le même que celui qui aurait été obtenu directement par 
l'application 
globale de la méthode. 

Cependant, la situation est un peu différente pour les RF, qui ne sont pas 
basées sur des
sommes. Comme indiqué dans \cite{delrio_etal_IS2014}, la version standard MR
des RF, noté MR-RF dans la suite, repose sur la construction parallèle
des sous-forêts d'arbres obtenus sur des sous-échantillons bootstrap des sous-ensembles
de données propres à chaque tâche Map : chaque bloc
de données est ainsi envoyé à un travail Map dans lequel une RF (qui peut avoir
un nombre modéré d'arbres) est construit (la clé de sortie est toujours égale à
1 et la valeur de sortie est la forêt). Il n'y a pas à proprement parler d'étape 
Reduce car la
sortie des travaux Map est une collection de sous-forêts qui, fusionnées,
donne la forêt finale. La prédiction elle-même, pour toutes les données, exige
un autre passage de MR, comme expliqué dans \cite{delrio_etal_IS2014}. Cette 
méthode
est celle mise en oeuvre dans la librairie Apache\(^\textrm{\tiny TM}\) Mahout.

Mentionnons enfin que dans ce contexte, la moyenne des erreurs OOB 
des sous-forêts est directement accessible et est en général, de bonne qualité. 
Elle peut néanmoins ne pas être proche de l'erreur OOB de la forêt globale si l'échantillonnage 
est trop hétérogène entre les différentes tâches Map. Une remarque semblable 
quant à l'importance des variables s'en déduit. Dans cette situation on peut employer
diverses stratégies pour se prévenir de ce risque (voir le paragraphe \ref{echantillonnage-et-blb} ainsi que \cite{genu-2015})).

\subsection{Forêts aléatoires en
ligne}\label{forets-aleatoires-en-ligne}

L'idée générale des RF en ligne (ORF dans la suite), introduite par 
\cite{saffari_etal_ICCV2009}, est d'adapter la méthodologie RF
au cas où les données arrivent séquentiellement, également connu sous le nom
flux de données. Ce cadre suppose qu'à l'instant \(t\) on ne peut 
avoir accès à toutes les données du passé, mais seulement à 
l'observation courante ou quelques observations récentes. Les
ORF ont été d'abord définies dans \cite{saffari_etal_ICCV2009}
et détaillées seulement pour la classification. Elles combinent l'idée 
du bagging en ligne de \cite{oza_russell_ICSMC2005}, les  
Extremely Randomized Trees (ERT) de \cite{Geurts06} 
et un mécanisme pour mettre à jour la forêt à chaque
instant d'arrivée d'une nouvelle observation.

Plus précisément, lorsqu'une nouvelle donnée arrive, le bagging en ligne
met à jour \(k\) fois un arbre donné, où \(k\)  est tiré au hasard suivant une 
loi
de Poisson pour imiter un échantillon bootstrap par lots. Cela signifie que 
cette 
nouvelle donnée apparaîtra \(k\) fois dans l'arbre.
Les forêts ERT sont utilisées au lieu des RF originales de Breiman, car 
elles permettent une mise à jour plus rapide : \(S\) découpes
 (définies par une variable et une valeur) sont tirées au hasard
pour chaque n\oe{}ud et le split final est optimisé uniquement parmi ces  \(S\)
splits candidats. En outre, toutes les décisions rendues par un arbre sont 
basées uniquement
sur les proportions de chaque label dans les observations dans un nœud.  ORF
maintient à jour une mesure d'hétérogénéité sur la base de
ces proportions, utilisées pour déterminer le label étiquetant un n\oe{}ud. Donc 
quand
un nœud est créé,  \(S\) splits candidats (donc \(2S\) nouveaux nœuds 
candidats) 
sont tirés au hasard et quand une nouvelle donnée arrive dans un n\oe{}ud 
existant, 
cette mesure est mise à jour pour tous les \(2S\) n\oe{}uds candidats. Ce
 mécanisme est répété jusqu'à ce qu'une condition d'arrêt soit réalisée et le
split final minimise la mesure d'hétérogénéité parmi les  \(S\)
splits candidats. Ensuite, un nouveau n\oe{}ud est créé et ainsi de suite.

L'erreur OOB d'un arbre \(t\) est aussi estimée en ligne : la donnée courante
est OOB pour tous les arbres pour lesquels la variable aléatoire de Poisson 
utilisée pour
répliquer l'observation dans l'arbre est égale à 0. La prédiction
fournie pour un tel arbre \(t\) est utilisée pour mettre à jour la valeur courante
de \(\mbox{errTree}_t\) l'erreur d'un arbre \(t\).
Cependant, la prédiction ne peut être réévaluée après que l'arbre ait
été mis à jour avec la prochaine donnée, cette approche n'est donc qu'une 
approximation de
\(\mbox{errTree}_t\) originale.

\subsection{Echantillonnage et BLB}\label{echantillonnage-et-blb}

L'une des limites importantes de MR-RF est le biais d'échantillonnage possible
provenant du découpage des données en plusieurs morceaux pour les travaux Map
exécutés en parallèle. Cela pourrait être pris en compte en améliorant le plan 
de
partition, en utilisant au moins une partition au hasard des données ou, 
probablement
plus efficacement, grâce à une stratification selon la variable  \(Y\), ou 
encore 
de sous-échantillonner plus soigneusement en exploitant le fait que l'ensemble 
des données n'est probablement pas nécessaire pour obtenir des estimations 
précises.

Une autre alternative intéressante à un sous-échantillonnage et la 
stratification est le
BLB (Bag of Little Bootstrap) décrite dans
\cite{kleiner_etal_JRSSB2014}. Cette méthode vise la construction 
d'échantillons bootstrap de taille \(n\), chacun contenant seulement \(m \ll n\) 
données différentes. La taille de l'échantillon est classique (\(n\)),
évitant ainsi le problème du biais impliqué par la méthode du \(m\) parmi \(n\) 
bootstrap, 
conséquence directe de la stratégie de MR
(chaque fragment contient une partie de l'ensemble de données et entraîne ainsi
un échantillon bootstrap de taille \(m\)). Le traitement de cet échantillon est
simplifié par une pondération intelligente et est donc gérable même pour
 \(n\) très grand puisqu'il ne contient qu'une petite fraction (\(m/n\)) 
d'observations 
 différentes de l'ensemble de données d'origine. Cette
approche est de surcroît, bien étayée par des résultats théoriques puisque les 
auteurs prouvent son équivalence avec la méthode bootstrap standard.

\subsection{Garantie théorique}\label{garantie-theorique-3}

En dehors du résultat de consistance de \cite{kleiner_etal_JRSSB2014}, mais qui
porte sur de l'estimation par bootstrap et n'est pas directement
applicable aux forêts BLB introduite précédemment, très peu de garanties
théoriques sont disponibles pour les variante big data de forêts aléatoires.

Une exception notable est le résultat de consistance pour des forêts
aléatoires en ligne, dû à \cite{denil_etal_ICML2013}. Les auteurs
introduisent une variante de ORF comportant deux principales différences.
Tout d'abord, aucun bootstrap en ligne n'est effectué et d'autre part, le flux de données est divisé au
hasard en deux flux : le flux de structure et le flux d'estimation. Les données
du flux de structure sont utilisées seulement pour l'optimisation des découpes,
tandis que les données du flux d'estimation ne sont utilisées que pour
attribuer un label à un n\oe{}ud. Rappelons que dans les ORF, et aussi dans
la variante de \cite{denil_etal_ICML2013}, les points de coupures sont
choisis avec une loi uniforme. Ceci permet des mises à jour très rapides,
nécessaires dans un cadre en ligne, mais permet également de simplifier
l'analyse théorique de la méthode.

De plus, le découplage en données de structure et données d'estimation,
qui permet d'avoir indépendance entre les labels associés aux
feuilles d'un arbre et la partition de l'arbre, facilite l'obtention du
résultat. Ce découplage, qui a déjà été introduit par \cite{biau_JMLR2012}
et utilisé dans des études de simulations par \cite{arlot2014analysis} et
\cite{arlot2016comments}, a également permis à ces mêmes auteurs d'obtenir
un résultat de consistance pour une variante (hors ligne) très proche des RF-RI
dans un cadre de régression dans \cite{denil2014narrowing}.

\section{Remerciements}
Nous tenons à remercier les collègues qui ont accompagné nos réflexions sur ces 
sujets au travers de nombreuses collaborations, en particulier Sylvain Arlot, 
Servane Gey, Christine Tuleau-Malot et Nathalie Villa-Vialaneix.

Signalons enfin que ce document est une version préliminaire d'un chapitre d'un ouvrage en préparation 
édité par Myriam Maumy-Bertrand, Gilbert Saporta et Christine Thomas-Agnan, 
Apprentissage Statistique et Données Massives, Technip, 2017, issu des 
Journées d'Etudes en Statistique de la SFdS, 2-7 octobre 2017, Fréjus, France.

\bibliographystyle{apalike}
\bibliography{jes_rf}

\end{document}